\documentclass[authoryear,final,3p]{elsarticle}

\usepackage{amssymb}
\usepackage{multirow}
\usepackage{url}
\usepackage{amsfonts}
\usepackage{amsmath}
\usepackage{float}
\usepackage{graphicx}

\newcommand{\FloatBarrier}{\par\vspace{0pt}}
\usepackage{svg}
\usepackage{array}
\usepackage{longtable}
\usepackage{subcaption}
\usepackage{booktabs}
\usepackage{xcolor}
\usepackage{colortbl}
\usepackage{pdflscape}

\journal{Transportation Research Part D: Transport and Environment}

\begin{document}

\begin{frontmatter}



\title{From Causal Discovery to Implementation: An Agentic AI Framework for E-Scooter Mobility Hub Planning Across 29 German Cities}

\author[label1,label2]{Meng Jin}
\author[label1]{Melanie Handrich}
\author[label1]{Simone Martinenz}
\author[label1]{Nicholas Hoeser}
\author[label2]{Ziyue Li\corref{cor1}}
\cortext[cor1]{Corresponding author. Email Address: ziyue.li@tum.de}

\affiliation[label1]{organization={Fraunhofer Institute for Industrial Engineering IAO},
            city={Heilbronn},
            country={Germany}}

\affiliation[label2]{organization={Department of Operations \& Technology, Heilbronn Data Science Center, Munich Data Science Institute, Technical University of Munich}, country={Germany}}
            
\begin{abstract}

Existing approaches to e-scooter mobility hub planning lack city-type-specific causal
evidence: demand models are typically correlational, built on proprietary trip data,
and do not distinguish how driver profiles vary across urban typologies.
This paper presents a three-phase agentic AI framework that constructs a
\textit{Causal Template Library} from public GBFS data across 29 German cities,
encoding which environmental features causally drive hotspot demand for each combination
of city type (large, university, industrial, hilly) and cluster type (core, peripheral).
A large language model (LLM)-orchestrated causal discovery pipeline adapts algorithm selection to local data
conditions across 57 city-cluster units.
The library reveals systematic variation: core demand is driven by activity access and
transit proximity, while peripheral demand responds to built form, with city-type-specific
patterns supporting transferable siting templates.
A planning tool built on the library scores candidate sites, calibrates infrastructure
recommendations to local demographics, and generates practitioner-ready reports.
In Heilbronn, Germany, two hub sites informed by the framework's causal evidence are currently
under construction, illustrating how the outputs can support real-world siting decisions.

\end{abstract}


\begin{graphicalabstract}
\includegraphics[width=\textwidth]{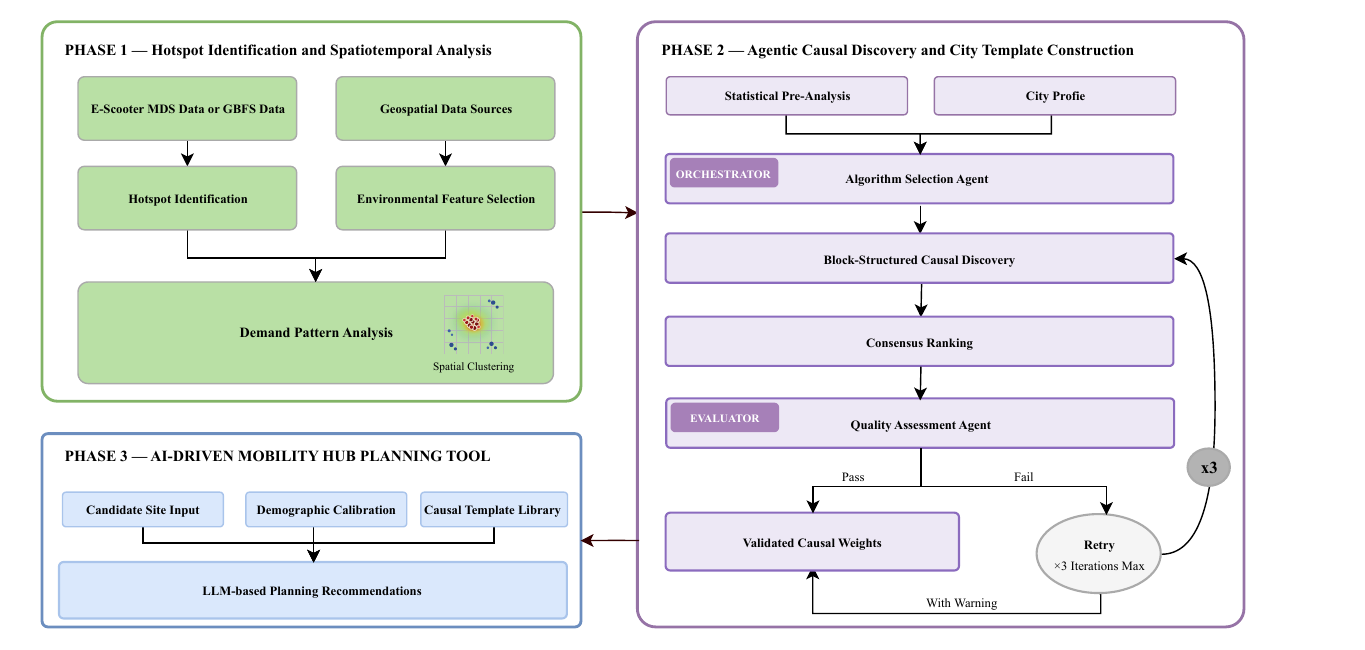}
\end{graphicalabstract}

\begin{highlights}
\item Three-phase agentic AI framework for e-scooter hub planning across 29 German cities
\item GBFS-only hotspot inference removes dependency on proprietary operator data
\item LLM-orchestrated causal discovery builds a transferable Causal Template Library
\item Facility recommendations calibrated to local demographics via preference survey
\item Framework validated in Heilbronn with two hubs currently under construction
\end{highlights}

\end{frontmatter}



\section{Introduction}
\label{sec1}

Shared e-scooter services gained federal regulatory approval in Germany in 2019 and expanded
rapidly in the years that followed \citep{guo2021understanding,badia2023shared}.
Without designated return points, scooters cluster on pavements and around transit stops
in ways that invite regulatory intervention.
City governments have increasingly adopted the \emph{mobility hub} as a structural response,
a fixed node that concentrates parking, charging, and connections to buses, trams, and shared
bicycles.
The question has shifted from whether to build such infrastructure to \emph{where} to site it
and \emph{what} to put in it.

Answering that question rigorously requires city-type-specific causal knowledge that does
not yet exist.
Existing demand studies identify associations between e-scooter ridership and urban features
such as transit density, cycling infrastructure, and land use mix
\citep{caspi2020spatial,huo2021influence,mckenzie2019spatiotemporal}, but correlational evidence
is insufficient for capital investment decisions.
Hub construction ties up resources for years and is hard to reverse: planners need to know
which built-environment conditions \emph{produce} demand, not merely co-occur with it.
More critically, causal relationships are unlikely to be uniform across cities.
A large city anchored by major rail terminals has a fundamentally different demand structure
than a university city, a hilly city, or an industrial city
\citep{li2022comprehensive,li2024integrating,jiao2024spatial}.
Planning on pooled correlations risks misallocating resources in cities whose demand profile
deviates from the average.
What is needed is a structured library of city-type-specific causal demand templates that
practitioners can match to their context without re-estimating from scratch.

Three further obstacles stand between that goal and practical use.
First, constructing such a library at scale requires trip data that most cities cannot access:
operator records are shared only under bilateral agreements.
The General Bikeshare Feed Specification \citep[GBFS;][]{MobilityData} offers an open alternative, but its
suitability for planning-grade causal analysis has not been demonstrated.
Second, causal discovery methods must be adapted to the heterogeneous data conditions found
across dozens of cities, a task that is prohibitively labour-intensive without automation.
Third, even a well-grounded siting recommendation leaves open what facilities to install.
Facility priorities differ markedly by age group \citep{budnitz2025understanding,kazemzadeh2024modal}, and a hub designed without
reference to local demographics will miss part of its user base.

This paper presents a three-phase agentic AI framework that addresses these gaps, applied
across 29 German cities spanning four functional typologies and three population-size classes.
Its central output is a \emph{Causal Template Library}: a structured knowledge base encoding
which environmental features causally drive e-scooter hotspot demand for each combination of
city type (large, university, industrial, hilly) and cluster type (core, peripheral).
Phase~1 infers demand hotspots from public GBFS data and classifies them by local mobility
context.
Phase~2 runs an LLM-orchestrated causal discovery pipeline that autonomously adapts algorithm
selection to each city's data conditions and aggregates results into the library.
Phase~3 scores candidate sites against the matched library entry and calibrates facility
recommendations to local demographics using a nationally representative citizen preference
survey.
Heilbronn serves as the primary case study, validated against 12 months of proprietary
operator data. The same pipeline is applied to 28 further cities using only public GBFS.
Two hub sites in Heilbronn informed by the framework's causal evidence are currently under
construction.

The main contributions are:
\begin{itemize}
  \item A \emph{Causal Template Library} of city-type-specific demand drivers across
        29 German cities, providing transferable siting templates that generalise
        without city-specific data agreements.
  \item A demonstration that public GBFS data can replace operator trip records for
        planning-grade hotspot detection and causal analysis.
  \item An LLM-orchestrated causal discovery pipeline that scales across heterogeneous
        urban data conditions without manual reconfiguration.
  \item A planning tool integrating the library with age-stratified demographic
        calibration, deployed to all German municipalities.
\end{itemize}

The rest of the paper is organised as follows.
Section~\ref{sec2} reviews related work.
Section~\ref{sec3} describes the study areas and data sources.
Section~\ref{sec4} presents the three-phase methodology.
Section~\ref{sec5} presents results for Heilbronn and the 28 further cities.
Section~\ref{sec6} discusses findings and implications.
Section~\ref{sec9} summarises the conclusions.

\section{Related Work}
\label{sec2}

\subsection{E-Scooter Demand and Environmental Factors}

Trip origins and destinations concentrate around transit stations, universities, and
commercial districts, with transit density, cycling infrastructure, land use mix,
and POI concentration as consistent positive predictors and steep terrain as a
suppressant \citep{mckenzie2019spatiotemporal, caspi2020spatial, huo2021influence,
guo2021understanding}.
Comparative analyses across European cities confirm these patterns beyond North American
contexts while revealing systematic differences by city size and transit quality
\citep{li2022comprehensive, badia2023shared}.
Large-scale trip data from Munich further document demand rhythms tied to commuting and
leisure, with hotspots concentrating at rail interchange nodes
\citep{sellaouti2024munich}.
Weather exerts a consistent influence: temperature raises ridership and precipitation
suppresses it, with stronger seasonality in colder climates \citep{morton2025weather}.

These associations are spatially non-stationary.
Built-environment effects that are significant in urban cores can be negligible or
reversed at the urban fringe \citep{baijiao2020, jiao2024spatial}, and Bayesian
modelling confirms non-linear, typology-dependent effects on zonal demand
\citep{zhai2025bayesian}.
On the policy side, parking regulation affects ridership through zone density and
walking-distance penalties \citep{bergwincent2025parking}, e-scooters function as
transit feeders in high-coverage areas but substitute for transit where service is
weak \citep{jayawardhena2025ptlinks, li2024integrating}, and usage is strongly
age-graded and concentrated in higher-income areas \citep{budnitz2025understanding}.

Despite this evidence base, the dominant paradigm remains correlational.
Existing models identify statistical associations but cannot establish which
built-environment conditions \emph{produce} demand, a distinction that matters for
capital investment decisions \citep{caspi2020spatial, huo2021influence}.
No study has produced a city-type-specific causal demand library applicable across
diverse urban typologies without city-specific trip data.

\subsection{Causal Discovery in Urban Mobility}

Causal discovery algorithms, including PC \citep{spirtes2000causation}, GES
\citep{chickering2002optimal}, LiNGAM \citep{shimizu2006lingam}, NOTEARS
\citep{zheng2018notears}, and DAGMA \citep{bello2022dagma}, recover directed acyclic
graph structures from observational data and are increasingly applied to urban systems.
\citet{feng2025urban} used reinforcement learning to discover causal graphs among citizens,
locations, and mobility behaviours, showing that naive regression conflates association
with causation in ways that matter for planning.
This distinction has been quantified in public transport: a PC-algorithm study on Hong Kong
MTR smart card data found that the causal effect of a fare discount is less than
a third of the correlational estimate \citep{farecausal2024}.
Beyond mobility demand, causal structure learning has informed traffic prediction
\citep{causalgrit2025, lan2023mm, gou2024traffident}, EV charging station placement \citep{junker2025ev}, and urban
policy evaluation \citep{wang2025ulez, panama2025bn}.

Causal discovery has not been applied to e-scooter demand modelling across
heterogeneous cities, and no study has examined how causal driver profiles vary by city
type, the precondition for a transferable planning template library.

\subsection{AI-Assisted Mobility Planning Tools}

AI in urban mobility planning has evolved from demand forecasting and station-siting
decision support \citep{shulajkovska2024ai} towards active plan-making, in which systems
propose, evaluate, and iterate on spatial configurations \citep{peng2024pathway}.
Large language models (LLMs) have accelerated this shift: their capacity to integrate
heterogeneous data, reason across steps, and produce human-readable output makes them
well-suited to translating analytical evidence into planning recommendations
\citep{zheng2025llm}.
Agentic AI systems, in which LLM orchestrators autonomously decompose tasks and select
tools, represent the current frontier \citep{tiwari2025agentic}.

No existing system integrates causal demand evidence with demographic calibration and
LLM-generated planning reports into a deployable hub-siting tool applicable across
city types at national scale.

\FloatBarrier
\section{Study Area and Data}
\label{sec3}

\subsection{Study Area}

This study covers 29 German cities where shared e-scooter services were actively operated during the data collection period. The 29 cities are classified along two dimensions (Table~\ref{tab:cities}).
By population size, the sample comprises 6 large cities ($>$500K),
19 medium cities (100K--500K), and 4 small cities ($<$100K).
By functional character, four typologies capture urban characteristics
relevant to e-scooter demand patterns, and typologies are not mutually exclusive:
\emph{university cities} (4), \emph{port cities} (3),
\emph{hilly cities} (4), and \emph{industrial cities} (3).
The functional typology counts do not sum to 29,
as cities may appear in more than one category
and cities without a distinctive functional character are not assigned to any typology.
Jena is the only city assigned to two functional typologies,
classified as both a university city and a hilly city.

\begin{table}[ht]
\centering
\caption{Classification of 29 study cities by population size and functional typology.}
\label{tab:cities}
\small
\begin{tabular}{lcp{8cm}}
\hline
\textbf{Type} & \textbf{$n$} & \textbf{Cities} \\
\hline
\multicolumn{3}{l}{\textit{By population size}} \\
\quad Large ($>$500K)    & 6  & Berlin, Hamburg, K\"{o}ln, D\"{u}sseldorf, Dortmund, Stuttgart \\
\quad Medium (100--500K) & 19 & Bonn, Braunschweig, Chemnitz, Darmstadt, Erfurt,
                                Erlangen, Heilbronn, Ingolstadt, Jena, Kaiserslautern,
                                Karlsruhe, Kiel, Mannheim, Reutlingen, Rostock,
                                Saarbr\"{u}cken, Wolfsburg, Wuppertal, Pforzheim \\
\quad Small ($<$100K)    & 4  & B\"{o}blingen, Friedrichshafen, Ludwigsburg, T\"{u}bingen \\
\hline
\multicolumn{3}{l}{\textit{By functional typology (non-exclusive)}} \\
\quad University & 4 & Darmstadt, Erlangen, Jena,
                    T\"{u}bingen \\
\quad Port       & 3 & Hamburg, Kiel, Rostock \\
\quad Hilly      & 4 & Jena, Saarbr\"{u}cken, Stuttgart, Wuppertal \\
\quad Industrial & 3 & Chemnitz, Dortmund, Wolfsburg \\
\hline
\end{tabular}
\end{table}

\begin{figure}[htbp]
\centering
\begin{subfigure}[t]{0.48\textwidth}
  \centering
  \includegraphics[width=\textwidth, keepaspectratio]{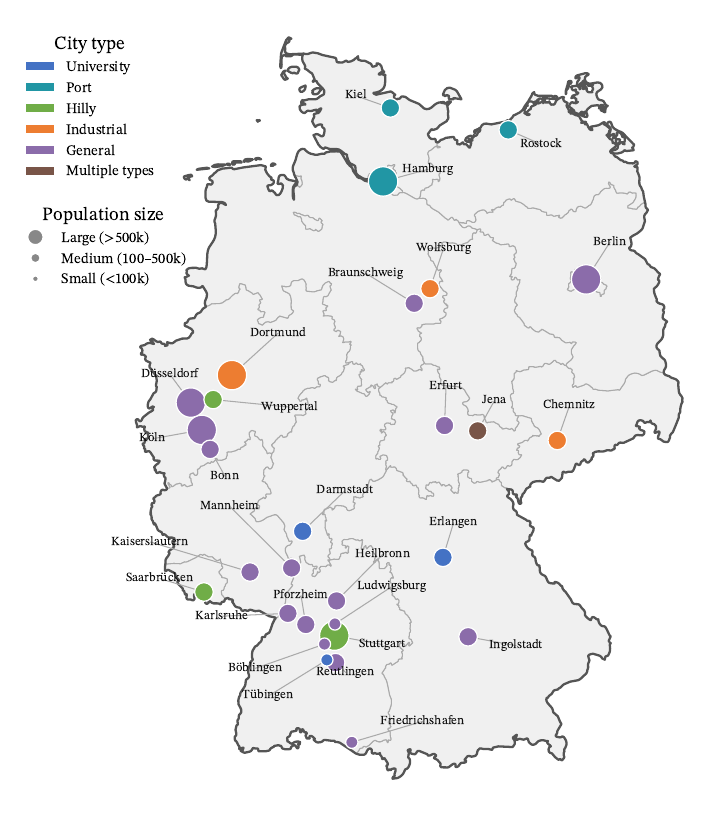}
  \caption{Geographic distribution of the 29 study cities across Germany.}
  \label{fig:study_area_map}
\end{subfigure}
\hfill
\begin{subfigure}[t]{0.48\textwidth}
  \centering
  \includegraphics[width=\textwidth, keepaspectratio]{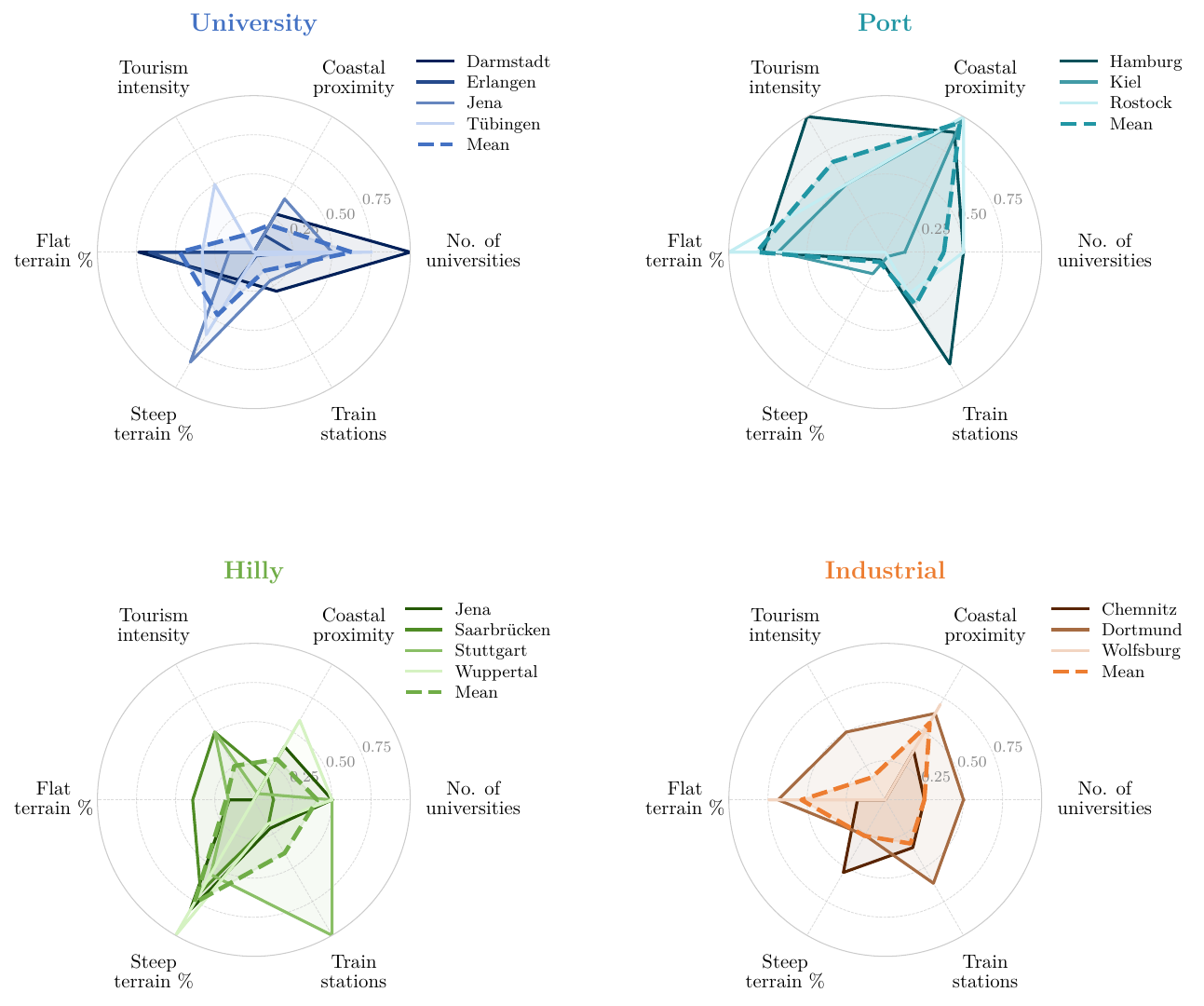}
  \caption{Environmental feature profiles by city functional typology.}
  \label{fig:study_area_radar}
\end{subfigure}
\caption{Study area overview.}
\label{fig:study_area}
\end{figure}

Figure~\ref{fig:study_area_map} shows the geographic distribution of the 29 study cities,
with point size reflecting population class and colour indicating functional typology.
Figure~\ref{fig:study_area_radar} shows normalised mean environmental feature profiles
for each functional typology, revealing distinct characteristics
that reflect the urban identity of each city type.
University cities are distinguished by a high number of universities, reflecting their
official designation as \emph{Universitätsstädte} (university towns) in Germany.
Port cities score highly on tourism-related features and coastal proximity,
and also exhibit notably flat terrain. Hamburg additionally records high train station density,
expected for a large city with a major rail interchange.
Hilly cities are defined above all by steep terrain,
with Stuttgart again recording high train station density as a large city.

\subsection{Primary Dataset: Heilbronn Operator Data}

Heilbronn serves as the primary case study: a medium-sized city of approximately 130,000 residents in Baden-Württemberg and an active municipal partner in the development and validation of the framework. The primary dataset comprises 12 months of trip-level data provided directly by Dott, covering approximately 379,000 e-scooter trips conforming to the Mobility Data Specification \citep[MDS;][]{OMF2019MDS} version 0.4, with full GPS route traces and millisecond-precision timestamps.

\subsection{Scalability Dataset: GBFS Open Data}

GPS position snapshots were collected from publicly available GBFS feeds across all 29 study cities between 10 April and 20 May 2026 (41 days), ingesting feeds from multiple operators (Dott, Bolt, Voi, Lime) through the same pipeline. Trip origins and destinations were inferred from consecutive position snapshots following \citep{Micromobility}.

\begin{figure}[htbp]
\centering
\includegraphics[width=\linewidth]{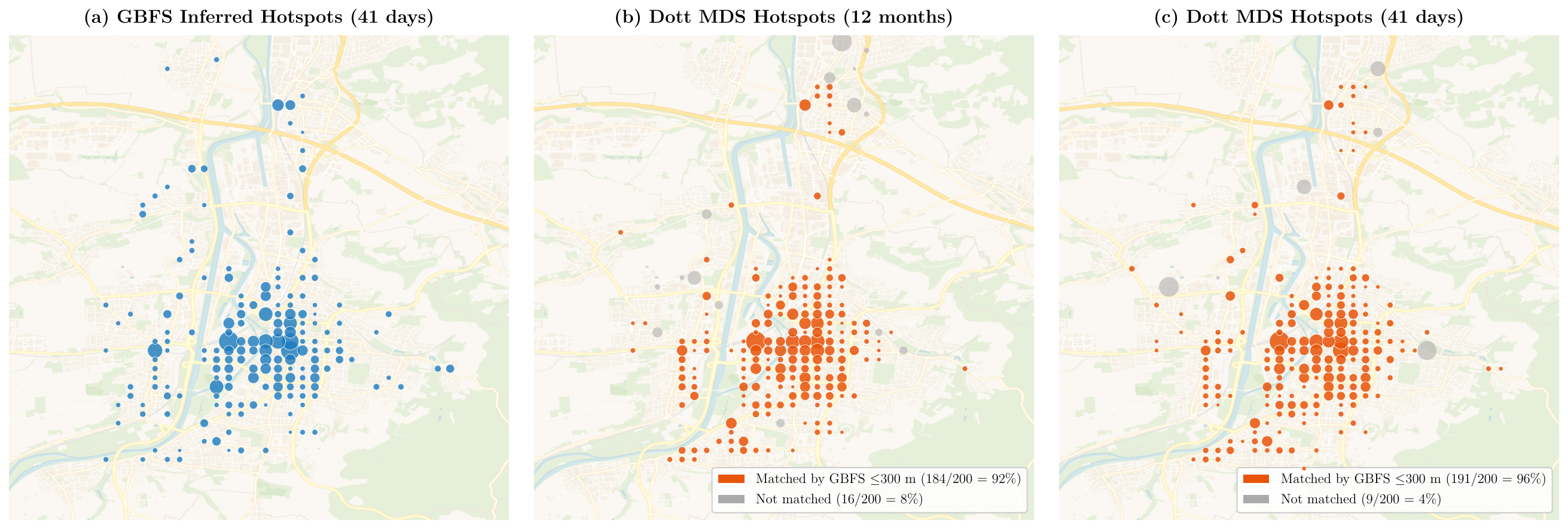}
\caption{Spatial validation of GBFS-derived hotspots against concurrent 41-day and full-year MDS data in Heilbronn (300\,m threshold).}
\label{fig:heilbronn_comparison}
\end{figure}

Validation in Heilbronn against a one-year proprietary MDS dataset (Dott, April 2025--April 2026, 378,857 trips) confirms high fidelity: 184 of 200 reference hotspots (92\%, $r = 0.869$) were recovered from the full-year comparison, rising to 191 of 200 (95.5\%, $r = 0.928$) when restricted to the same 41-day window (Figure~\ref{fig:heilbronn_comparison}). This establishes that 41 days of open GBFS data can replace proprietary trip records for hotspot detection.

Across all 29 cities, the pipeline yielded approximately 8.0 million inferred trips,
ranging from 5,496 in T\"{u}bingen to 1,847,147 in Stuttgart.
Origin and destination endpoints cluster spatially (mean overlap 85\%),
so hotspot detection merges both endpoint types before clustering.
The entire collection required no operator agreements,
making the approach applicable wherever GBFS feeds are publicly available.
\subsection{Environmental Features}

Environmental features are derived from five open geospatial data sources.
The majority of features, including public transport stops and routes, cycling infrastructure,
points of interest, road network, building footprints, parking facilities,
student dormitories, and parcel lockers, are extracted from
OpenStreetMap (OSM) via Geofabrik \citep{Geofabrik}. Population counts and age distributions are obtained from
the German Census at 100\,m grid resolution \citep{Zensus}. Building heights and high-rise counts are derived from the GlobalBuildingAtlas \citep{GlobalBuildingAtlas},
a global dataset of 2.75 billion LoD1 building models at 3\,m resolution. Terrain flatness is derived from NASA SRTM 1-arcsecond elevation tiles \citep{SRTM},
with road segments classified as flat when slope is below 5\%.
Car-sharing and e-scooter station locations are obtained from GBFS real-time feeds.
For each city, the MobiData-BW aggregator \citep{MobiDataBW} is queried first.
Where no feed is available, the MobilityData global systems directory \citep{MobilityData}
is used to look up city-specific feed URLs individually.

\begin{figure}[htbp]
\centering
\includegraphics[width=\linewidth, height=0.55\textheight, keepaspectratio]{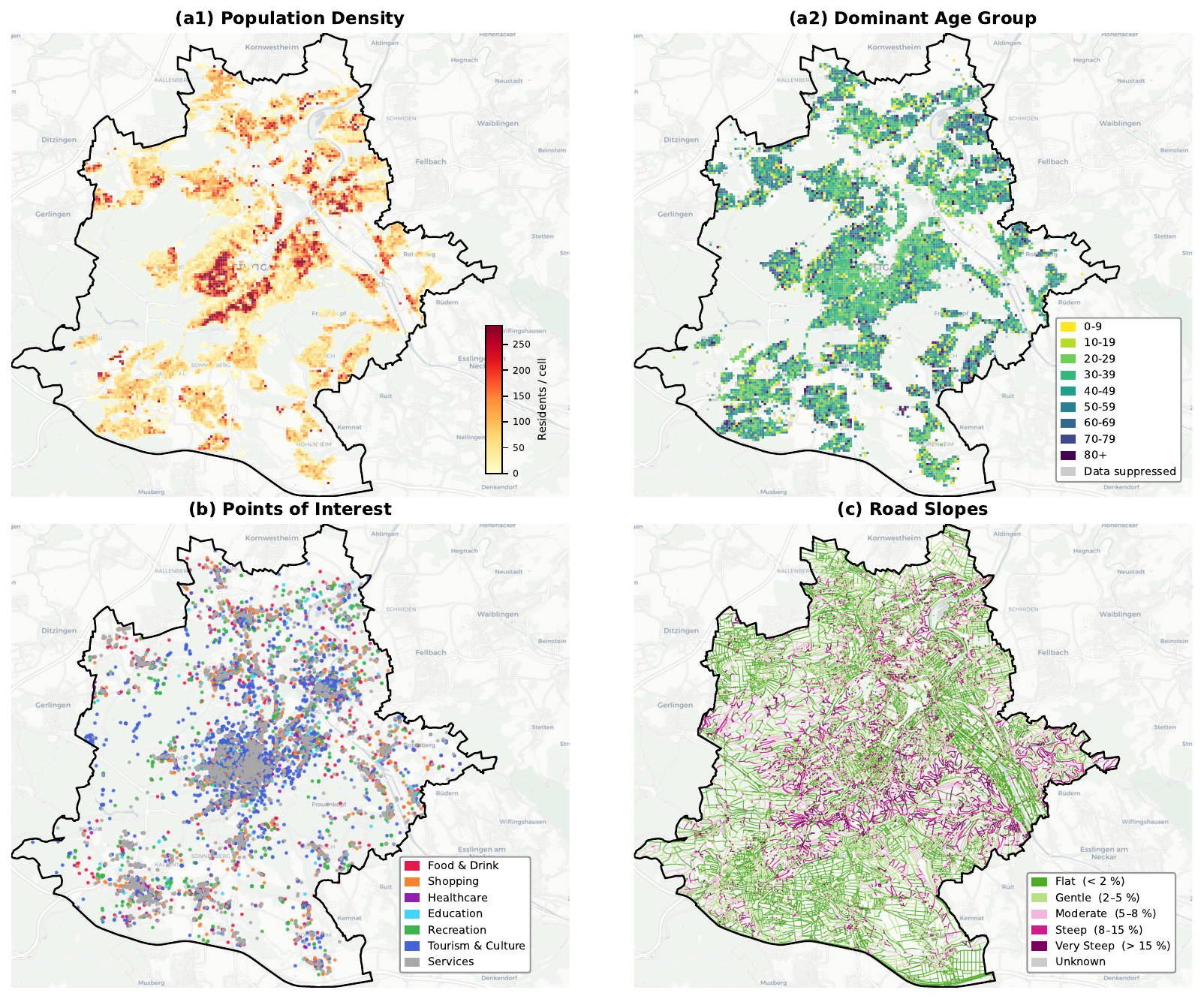}
\caption{Representative environmental feature layers for Stuttgart:
(a1)~population density at 100\,m grid resolution,
(a2)~dominant age group distribution,
(b)~POI distribution,
(c)~road slope classification.}
\label{fig:env_layer_stuttgart}
\end{figure}

All features are organised into five semantic blocks to mitigate multicollinearity
(Section~\ref{sec4}).
Table~\ref{tab:features} lists the full feature set with descriptions.
Features are derived from diffenrent geospatial layers, figure~\ref{fig:env_layer_stuttgart} illustrates four representative feature layers applied across all study cities, using Stuttgart as an example.

\small
\setlength{\tabcolsep}{4pt}
\renewcommand{\arraystretch}{1.0}
\begin{longtable}{p{2.5cm}p{3.8cm}p{6.2cm}}
\caption{Environmental feature engineering and selection}
\label{tab:features}\\
\hline
\shortstack[l]{\textbf{Feature}\\\textbf{Category}} & \textbf{Feature Name} & \textbf{Description} \\
\hline
\endfirsthead
\hline
\shortstack[l]{\textbf{Feature}\\\textbf{Category}} & \textbf{Feature Name} & \textbf{Description} \\
\hline
\endhead
\hline
\endfoot
\multirow{8}{*}{\textbf{Transportation}}
 & Bus stop count            & Number of bus stops within hotspot area \\
 & Train station count       & Number of train stations within hotspot area \\
 & Nearest bus stop (m)      & Distance to nearest bus stop \\
 & Nearest train station (m) & Distance to nearest train station \\
 & Bus route count           & Number of bus routes serving the hotspot \\
 & Train route count         & Number of train routes serving the hotspot \\
 & Cycling infra (km)        & Total length of cycling infrastructure \\
 & Slow roads (\%)           & Share of residential and pedestrian roads \\
\hline
\multirow{7}{*}{\textbf{POIs}}
 & POI education             & Count of educational facilities \\
 & POI food \& drink         & Count of food and drink venues \\
 & POI healthcare            & Count of healthcare facilities \\
 & POI recreation            & Count of recreational facilities \\
 & POI services              & Count of service establishments \\
 & POI shopping              & Count of retail outlets \\
 & POI tourism \& culture    & Count of tourism and cultural sites \\
\hline
\multirow{7}{*}{\textbf{Urban Form}}
 & Street intersections      & Number of street intersections \\
 & Residential buildings     & Count of residential buildings \\
 & Commercial buildings      & Count of commercial buildings \\
 & High-rise count           & Count of buildings exceeding 6 floors \\
 & Building coverage (\%)    & Fraction of hotspot area covered by buildings \\
 & Structured parking        & Count of structured parking facilities \\
 & Flat terrain (\%)         & Share of terrain with slope $<$5\% \\
\hline
\multirow{2}{*}{\textbf{Population}}
 & Population total          & Total resident population in catchment area \\
 & Dormitory count           & Number of student dormitories \\
\hline
\multirow{3}{*}{\shortstack[l]{\textbf{Mobility}\\\textbf{Supply}}}
 & Car-sharing stations      & Count of car-sharing stations \\
 & E-scooter stations        & Count of designated e-scooter parking stations \\
 & Parcel locker count       & Count of parcel lockers \\
\hline
\end{longtable}

\subsection{Citizen Preference Survey}

A citizen preference survey was conducted in May--June 2025 with a nationally
representative German sample ($n = 1{,}000$) and a Heilbronn-specific sample
($n = 298$ city, $n = 119$ region).
It covered hub usage rates, preferred locations, mobility mode integration, and
infrastructure priorities, using a five-point Likert scale normalised to 0--100\%
for analysis.

\begin{figure}[htbp]
  \centering
  \includegraphics[width=\linewidth, height=0.55\textheight, keepaspectratio]{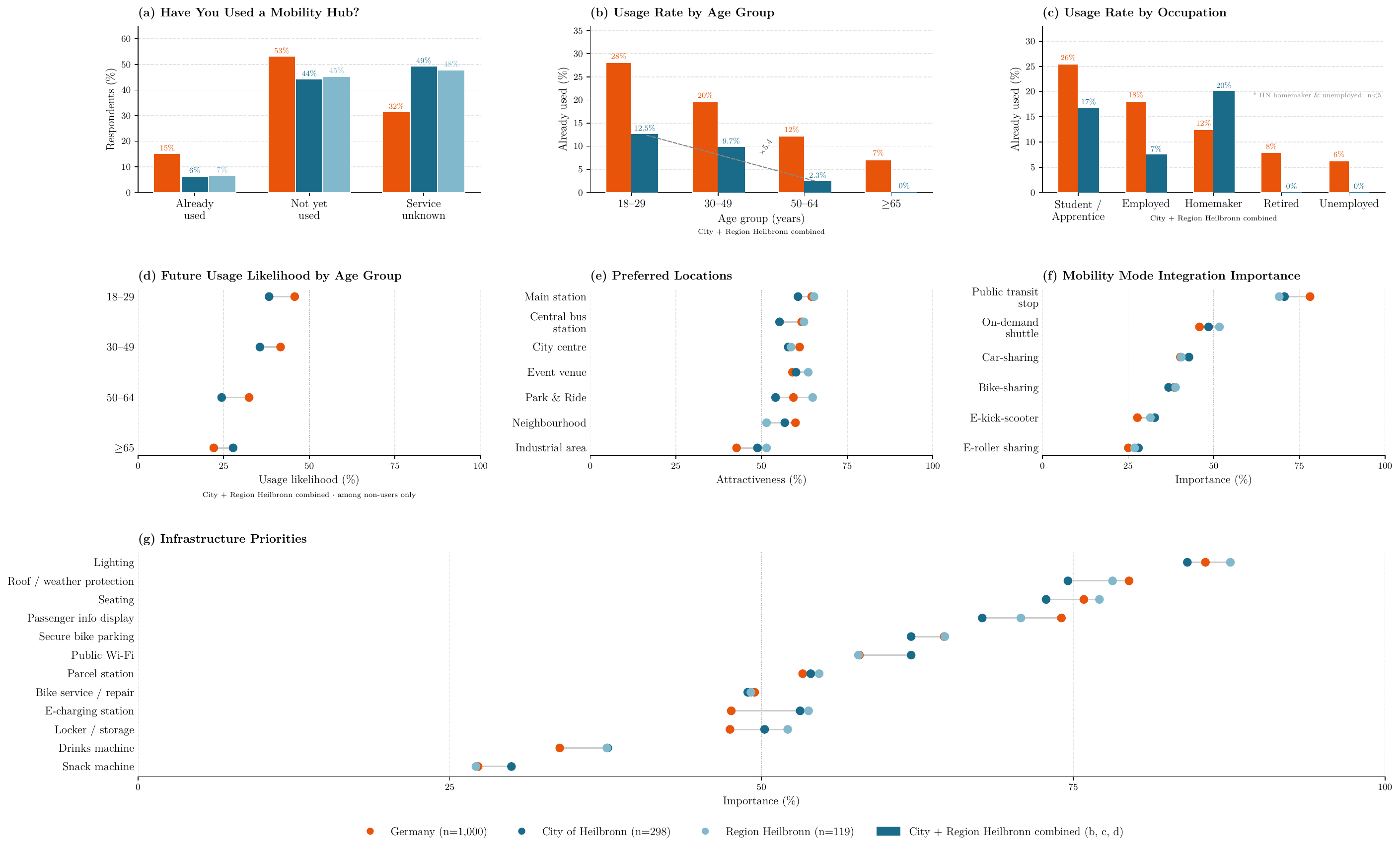}
  \caption{Citizen preference survey results: hub usage rates, location attractiveness, mobility mode importance, and infrastructure priorities across national, city, and regional Heilbronn samples (May--June 2025).}
  \label{fig:survey_key_findings}
  \end{figure}

Figure~\ref{fig:survey_key_findings} summarises the results across four dimensions.
Hub usage is strongly age-graded: nationally 15\% of respondents have used a mobility hub,
but usage among 18--29 year-olds is more than five times that of the 50--64 group in Heilbronn
(12.5\% vs.\ 2.3\%), and future usage likelihood declines from 46\% among the youngest
national cohort to 22\% among those aged 65 and above.
Main railway stations are the most preferred hub location across all three samples (61--65\%),
while industrial areas are consistently least attractive (43--52\%).
Public transport integration is rated the single most important mobility feature (78\%
nationally), far above car-sharing (40\%), bike-sharing (39\%), and micromobility options
(25--28\%).
Infrastructure priorities are consistent across groups: lighting (86\%), weather protection
(80\%), seating (76\%), and passenger information displays (74\%) all rank above the neutral
midpoint, while e-charging stations show the strongest regional variation (48\% nationally
vs.\ 53\% in Heilbronn).

These responses inform the demographic calibration in Phase~3: age-stratified usage probabilities and infrastructure preference weights are derived from the survey data and applied to each candidate site's catchment population.

\FloatBarrier
\section{Methodology}
\label{sec4}

\begin{figure}[htbp]
\centering
\includegraphics[width=\textwidth, height=0.82\textheight, keepaspectratio, clip]{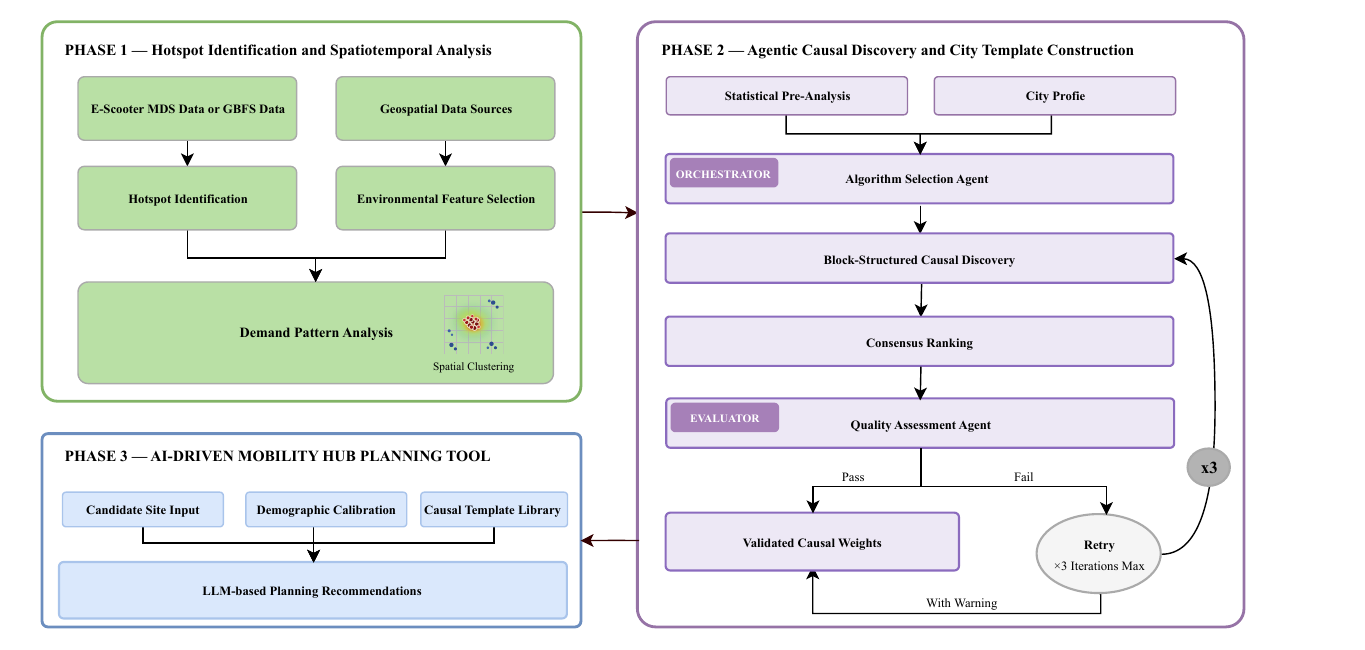}
\caption{Three-phase agentic AI framework for e-scooter mobility hub planning.}
\label{fig:framework}
\end{figure}

Figure~\ref{fig:framework} illustrates the three-phase framework. Phase~1 infers demand hotspots from MDS or GBFS trip data, calibrates the spatial clustering parameters, and partitions hotspots into environmental-feature-based clusters. Phase~2 runs an LLM-orchestrated causal discovery pipeline independently for each city-cluster unit, applying block-structured analysis to overcome global multicollinearity, and aggregates the results into a \textit{Causal Template Library} indexed by city typology and cluster type. Phase~3 is a practitioner-facing planning tool: candidate sites are scored against the matched causal weight vector, infrastructure recommendations are calibrated to the site's catchment age distribution, and an LLM synthesises the outputs into a structured planning report.

\subsection{Phase 1: Hotspot Identification and Spatiotemporal Analysis}

\subsubsection{Hotspot Identification}

Trip origins and destinations are obtained either directly from MDS operator records
or inferred from GBFS availability snapshots.
For each city, origin and destination endpoints are binned separately into 200\,m grid cells.
The top-$N$ cells accounting for 80\% of trips are retained and submitted to
DBSCAN \citep{khan2014dbscan} ($\varepsilon = 300$\,m, \texttt{min\_samples} $= 1$)
to merge adjacent high-demand cells into hotspots.
A trip-count-weighted centroid is computed for each cluster:
\begin{equation}
  \bar{\mathbf{x}}_k = \frac{\sum_{i \in C_k} w_i \mathbf{x}_i}{\sum_{i \in C_k} w_i}
  \label{eq:centroid}
\end{equation}
where $C_k$ is the set of grid cells in cluster $k$, $\mathbf{x}_i$ is the coordinate
of cell $i$, and $w_i$ is its trip count.
The procedure runs separately for origins and destinations, producing two hotspot sets per city.
A second DBSCAN pass merges the two sets, and hotspots within 300\,m of a counterpart
in the opposing set are flagged as bidirectional demand nodes.
Grid resolution (200\,m), coverage threshold (80\%), and clustering radius ($\varepsilon = 300$\,m)
were calibrated on the Heilbronn dataset, \ref{app:sensitivity} reports
the corresponding sensitivity analysis.

\subsubsection{Spatial Feature Extraction and Demand Pattern Analysis}

For each retained hotspot, a multi-modal isochrone of duration $t^*$ is computed via OSMnx \citep{boeing2017osmnx},
covering walking, cycling, e-scooter, and car-sharing access modes,
from which 27 environmental features (Table~\ref{tab:features}) are extracted.
Hotspots with degenerate isochrones (zero area) are excluded as noise.
The remaining hotspots within each city are clustered by their environmental feature vectors
using KMeans, with the optimal number of clusters selected automatically via the silhouette score.
Clusters are ranked by mean trip count and labelled $C_0, C_1, \ldots$ in descending order
of mobility intensity. Each city--cluster combination forms one independent analytical unit passed to the causal
discovery pipeline in Phase~2.

\subsection{Phase 2: Agentic Causal Discovery and City Template Construction}

\subsubsection{City Profile and Statistical Pre-Analysis}

Before entering the agentic loop, two structured JSON documents are prepared
for each city--cluster combination and passed directly to the Orchestrator.

\paragraph{City Profile}
Urban context modulates which environmental features are meaningful predictors of e-scooter demand. A city profile is assembled from three structured sources: Wikidata metadata (population, university count, city type), OpenStreetMap-derived statistics (terrain slope, building density, transit coverage), and behavioural summaries from the trip data (demand concentration, weekday-to-weekend ratio, peak-hour share). An LLM synthesis step condenses these into a concise city-profile JSON for the Orchestrator. Profiles were validated on five representative cities by two domain experts before full deployment.

\paragraph{Statistical Pre-Analysis}
The statistical pre-analysis quantifies the key properties of each cluster's feature matrix ($n$ hotspots, 27 features) to inform data-adaptive algorithm selection. The following indicators are computed and passed to the Orchestrator:

(i)~\textit{Basic structure}: sample size $n$, number of features $p$, and the ratio $n/p$.

(ii)~\textit{Distributional properties}.
Several causal discovery algorithms assume Gaussian-distributed inputs.
When this assumption is widely violated, non-Gaussian methods should be prioritised.
To quantify the extent of non-normality, the Shapiro--Wilk test \citep{gonzalez2019shapiro}
is applied to each feature, and the share of features that reject normality is recorded as:
\begin{equation}
  \rho_{\text{norm}} = \frac{1}{p}\sum_{j=1}^{p} \mathbf{1}[p_{\text{SW},j} < 0.05]
  \label{eq:rho_norm}
\end{equation}

(iii)~\textit{Target variable transformation}.
The Shapiro--Wilk test is applied separately to $y = \texttt{trip\_count}$ to assess
whether the target variable itself requires transformation.
Trip counts are typically right-skewed, with a small number of hotspots accounting for
a disproportionate share of demand.
When this skewness is severe, log-transforming $y$ stabilises regression-based causal estimates.
A flag $\delta_{\log}$ is set when both non-normality and strong skewness are confirmed:
\begin{equation}
  \delta_{\log} = \mathbf{1}\bigl[p_{\text{SW}}(y) < 0.05 \;\wedge\; \gamma_1(y) > 1.5\bigr]
  \label{eq:log_flag}
\end{equation}
where $p_{\text{SW}}(y) < 0.05$ follows the standard significance threshold for normality testing
and $\gamma_1(y) > 1.5$ follows the skewness criterion of \citep{box1964analysis}.

\subsubsection{Orchestrator: Algorithm Selection Agent}

Seven causal discovery algorithms spanning four algorithmic paradigms are available to the
Orchestrator (Table~\ref{tab:algorithms}).
Score-based methods (DAGMA, NOTEARS, GES) cast structure learning as continuous optimisation
over an acyclicity constraint \citep{zheng2018notears,bello2022dagma}.
DirectLiNGAM exploits non-Gaussianity for causal direction identification via ICA \citep{shimizu2011directlingam}.
Constraint-based methods (PC, FCI) use conditional independence tests but require sufficient
sample size relative to feature count \citep{kalisch2007estimating,spirtes2000causation}.
GRaSP uses a distribution-free permutation search, offering robustness when normality is
widely violated \citep{lam2022greedy}.

\begin{table}[htbp]
\caption{Causal discovery algorithms available to the Orchestrator, grouped by category.}
\label{tab:algorithms}
\centering
\small
\begin{tabular}{llp{6.2cm}}
\hline
\textbf{Category} & \textbf{Algorithm} & \textbf{Key assumptions / notes} \\
\hline
\multirow{3}{*}{Score-based}
  & DAGMA   & Acyclicity via log-determinant, continuous optimisation \citep{bello2022dagma} \\
  & NOTEARS & Acyclicity via trace exponential, linear SEM \citep{zheng2018notears} \\
  & GES     & Greedy equivalence search, Markov condition + faithfulness \citep{spirtes2000causation} \\
\hline
ICA-based
  & DirectLiNGAM & Non-Gaussian errors, linear acyclic SEM \citep{shimizu2011directlingam} \\
\hline
\multirow{2}{*}{Constraint-based}
  & PC  & Faithfulness, CI tests, requires sufficient $n/p$ \citep{kalisch2007estimating} \\
  & FCI & Faithfulness, allows latent confounders \citep{spirtes2000causation} \\
\hline
Permutation-based
  & GRaSP & Permutation search, robust under violations of faithfulness \citep{peters2017elements} \\
\hline
\end{tabular}
\end{table}

Phase 2 is implemented as an agentic loop in which an LLM Orchestrator (\texttt{nvidia/\allowbreak Orchestrator-8B}) selects
from these methods and configures hyperparameters for each city--cluster combination.
Rather than relying on the LLM's general reasoning, the Orchestrator's decisions are grounded in
explicit selection rules encoded in a system-prompt knowledge base,
each derived from the methodological literature \citep{kalisch2007estimating,shimizu2011directlingam,zou2005regularization,efron1993bootstrap},
making algorithm selection transparent, reproducible, and independently auditable.
Selection proceeds in two steps: rule-based exclusions first remove algorithms whose
assumptions are violated for the given unit, after which the Orchestrator autonomously
selects 3--4 methods from the remaining eligible set, reducing unnecessary computation.

The full candidate pool $\mathcal{S}_0$ comprises all seven algorithms in
Table~\ref{tab:algorithms}.
The active set for each city--cluster unit is obtained by excluding algorithms whose
data requirements are not met:
\begin{equation}
  \mathcal{S} = \mathcal{S}_0 \setminus \mathcal{S}_{\text{excl}}
  \label{eq:algset}
\end{equation}
The exclusion set $\mathcal{S}_{\text{excl}} = \mathcal{E}_1 \cup \mathcal{E}_2$
is the union of two independent conditions, all thresholds calibrated via sensitivity
analysis (\ref{app:n_threshold}-\ref{app:np_threshold}):
\begin{equation}
  \mathcal{E}_1 = \begin{cases}
    \{\text{GRaSP}\} & \text{if } n < n_{\min} \text{ or } \hat{\rho} \leq \rho_{\text{norm}} \\
    \emptyset        & \text{otherwise}
  \end{cases}
  \label{eq:e1}
\end{equation}
\begin{equation}
  \mathcal{E}_2 = \begin{cases}
    \{\text{PC, FCI, GES}\} & \text{if } n/p < \tau \\
    \emptyset                          & \text{otherwise}
  \end{cases}
  \label{eq:e2}
\end{equation}
where $n_{\min} = 50$, $\rho_{\text{norm}} = 0.5$, and $\tau = 5$.

The Orchestrator receives the city profile and statistical pre-analysis report as inputs.
Its output is a structured JSON plan that specifies the selected algorithm subset
and the minimum method-agreement count required to declare a high-confidence causal edge.
Global feature selection is not performed at this stage.
Global VIF exceeds 10 for 26 of the 27 environmental features, so any single-pool
selection step would either discard most features or retain a severely collinear set.
Instead, the 27 features are partitioned \textit{a priori} into five semantic blocks
(Section~\ref{sec3}), each containing features of low mutual collinearity,
making block-level causal inference tractable.

\subsubsection{Block-Structured Causal Discovery and Consensus Ranking}
\label{sec:block}

Causal discovery proceeds in two stages.
In Stage~1, each of the five semantic blocks defined in Table~\ref{tab:features}:
\textit{Transportation}, \textit{POIs}, \textit{Urban Form}, \textit{Population},
and \textit{Mobility Supply}, is analysed independently
with \texttt{trip\_count} as the target variable.
For every Orchestrator-selected algorithm, the raw causal graph is estimated,
and a bootstrap procedure re-runs the algorithm on $B$ resamples \citep{efron1993bootstrap}.
The stability of each edge $e$ is measured by its bootstrap frequency:
\begin{equation}
  f_{\text{boot}}(e) = \frac{1}{B}\sum_{b=1}^{B} \mathbf{1}\!\left[e \in \hat{G}^{(b)}\right]
  \label{eq:fboot}
\end{equation}
where $\hat{G}^{(b)}$ is the estimated causal graph on the $b$-th resample.
Edges appearing in more than half of the bootstrap resamples are retained as stable \citep{buhlmann2010variable}.
Consensus is computed across methods within each block using only directed edges.
\textit{High-confidence} direct causes require agreement among all selected methods,
while \textit{medium-confidence} causes require agreement among at least two methods. In Stage~2, direct causes of \texttt{trip\_count} are merged across all five blocks into into high- and medium-confidence sets. 

\subsubsection{Evaluator: Quality Assessment Agent}

The Evaluator agent (\texttt{MiniMaxAI/MiniMax-M2.5}) assesses each causal result along
three dimensions.

\textit{Bootstrap convergence} $\bar{f}_{\text{boot}}$ is the mean stability fraction
across all retained edges:
\begin{equation}
  \bar{f}_{\text{boot}} = \frac{1}{|\mathcal{C}|}\sum_{e \in \mathcal{C}} f_{\text{boot}}(e)
  \label{eq:fboot_mean}
\end{equation}
where $\mathcal{C}$ is the set of identified direct causes and $f_{\text{boot}}(e)$ is
defined in Eq.~(\ref{eq:fboot}).

\textit{Method consensus rate} $r_{\text{cons}}$ measures the mean fraction of selected
algorithms that agree on each edge:
\begin{equation}
  r_{\text{cons}} = \frac{1}{|\mathcal{C}|}\sum_{e \in \mathcal{C}}
    \frac{\bigl|\{m \in \mathcal{S} : e \in \hat{G}_m\}\bigr|}{|\mathcal{S}|}
  \label{eq:rcons}
\end{equation}
where $\hat{G}_m$ is the causal graph estimated by method $m$.

\textit{Domain consistency} $d \in \{0,1\}$ is a binary judgement by the Evaluator
on whether the identified causal drivers are plausible from an urban-transport-planning
perspective (e.g.\ \textit{bus stop density $\rightarrow$ e-scooter demand} is
domain-consistent, while \textit{building coverage $\rightarrow$ number of transit lines} is not).

A result is classified as \textit{pass} when $\bar{f}{\text{boot}}$ and $r{\text{cons}}$ are at acceptable levels, $d = 1$, and $|\mathcal{C}| > 0$; otherwise as \textit{partial} or \textit{fail} depending on how many conditions are violated. Failed results trigger an automated retry, passing the Evaluator's feedback to the Orchestrator as a revised prompt, up to three iterations.

\subsubsection{Causal Template Library Construction}

After all city--cluster units have been processed, their causal results are
aggregated into a \textit{Causal Template Library}.
Each record in the library stores: (i) the name of the causal driver feature,
(ii) a confidence tier (\textit{high} or \textit{medium}) derived from $\bar{f}_{\text{boot}}$
and $r_{\text{cons}}$, (iii) the cross-city occurrence frequency of that feature as a
direct cause of \texttt{trip\_count}, (iv) a stratified breakdown by city type
(large city, university city, industrial city, hilly city, etc.), and (v) the semantic
feature block to which the driver belongs.

\subsection{Phase 3: AI-Driven Mobility Hub Planning Tool}

\subsubsection{Candidate Site Input and Scoring}
Phase~3 is a practitioner-facing planning tool.
Planners place candidates on a map interface, the system computes multi-modal
accessibility isochrones (walking, cycling, public transit) and extracts each site's
urban profile from open geospatial sources described in Section~\ref{sec3}.
The open-data pipeline makes the tool applicable beyond the 29 study cities to any German municipality.
The planner selects the matching Causal Template Library entry by city typology and urban context type, the system retrieves the corresponding causal weight vector $\boldsymbol{\beta}$ and computes a composite suitability score for each candidate.

\begin{equation}
  S_c = \frac{\sum_{k} \beta_k \cdot x_{c,k}}{\sum_{k} |\beta_k|}
  \label{eq:score}
\end{equation}
where $x_{c,k}$ is the normalised feature value at site $c$ and the denominator
normalises $S_c$ to $[0,1]$.

\subsubsection{Demographic Calibration}

The tool integrates a citizen preference survey (full calibration tables in
\ref{app:survey}) to produce two site-level outputs.

\textbf{Mobility Usage Likelihood.}
Binary logistic regressions fitted on the survey data yield age-stratified usage
probabilities for four mobility modes, encoded as a
probability matrix $M \in \mathbb{R}^{4 \times 4}$ (age groups $\times$ modes).
The catchment age distribution $\pi_{c,a}$ is derived from the Census data intersected with the isochrone, giving the modal demand profile:
\begin{equation}
  \mathbf{m}_c = \sum_{a} \pi_{c,a} \cdot M_{a,:}
\end{equation}

\textbf{Companion Service Demand.}
Survey importance ratings for five hub facility categories are
stratified by age group and area type, forming a preference matrix
$W \in \mathbb{R}^{4 \times 5}$.
The age-weighted service demand vector for each site is:
\begin{equation}
  \mathbf{p}_c = \sum_{a} \pi_{c,a} \cdot W_{a,:}
\end{equation}
Both $\mathbf{m}_c$ and $\mathbf{p}_c$ are displayed in the dashboard alongside
city-level averages, translating demographic context into prioritised infrastructure
recommendations.

\subsubsection{LLM-based Planning Recommendations}

An LLM synthesises the candidate site profiles, suitability scores $S_c$,
and demographic demand vectors $\mathbf{m}_c$, $\mathbf{p}_c$ into a structured
planning report, with each feature benchmarked against the city-level distribution
to make site-specific deviations explicit.
The report covers four elements: a narrative \emph{site recommendation} naming the
top-ranked candidate with its principal causal rationale, a \emph{site ranking} of
all candidates by $S_c$, a \emph{site-level analysis} of strengths and planning
constraints for each candidate, and a \emph{companion service prioritisation} derived
from $\mathbf{p}_c$.
The output is written for municipal practitioners without prior knowledge of causal
inference methodology.

\FloatBarrier
\section{Results}
\label{sec5}

\subsection{Heilbronn: From Causal Discovery to Hub Construction}
\label{sec5a}

The framework is applied in full to Heilbronn using a 12-month operator dataset, serving as the primary proof-of-concept case.

\subsubsection{Phase 1: Hotspot Identification and Spatiotemporal Analysis}
The pipeline yields 200 hotspots covering 79\% of all 378,857 trip starts.
Start and end hotspot sets are spatially near-identical (Table~\ref{tab:overlap},
median separation 7\,m, Pearson $r = 0.972$) and are therefore treated as one unified set.

\begin{figure}[htbp]
\centering
\begin{subfigure}[b]{0.49\textwidth}
  \centering
  \includegraphics[width=\textwidth]{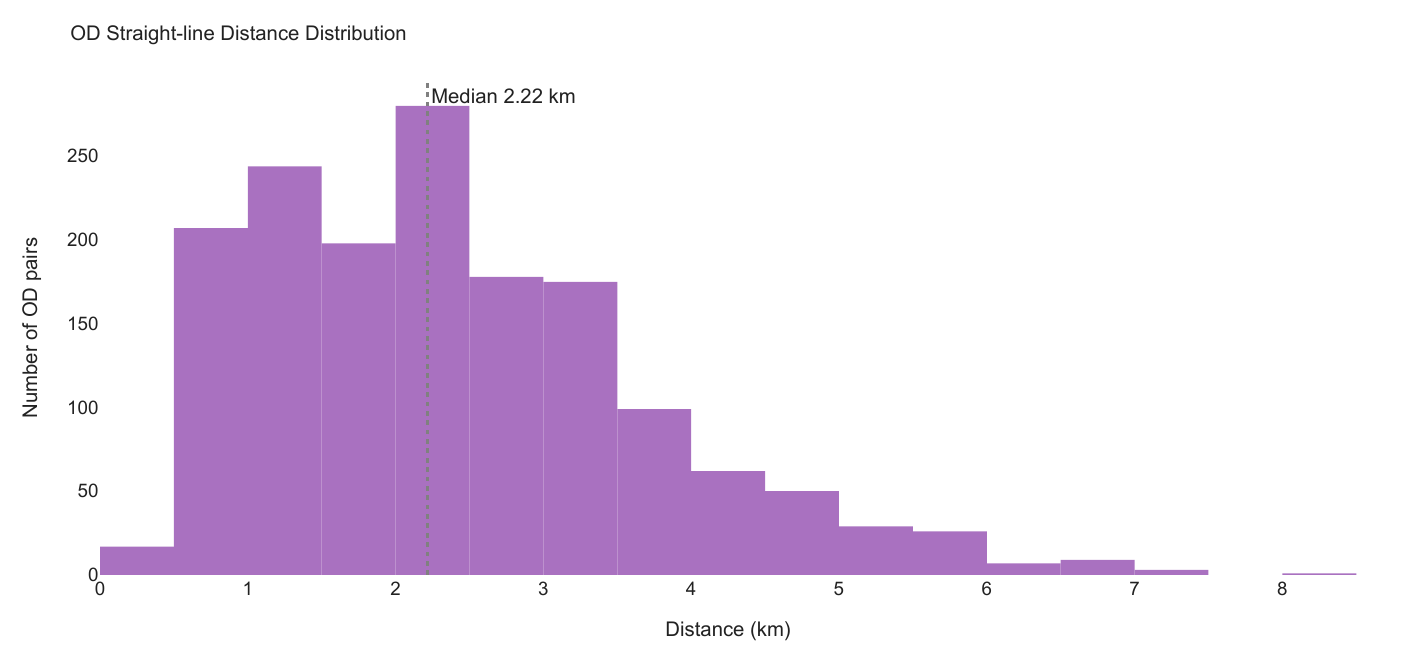}
  \caption{OD straight-line distance distribution.}
  \label{fig:od_dist}
\end{subfigure}
\hfill
\begin{subfigure}[b]{0.49\textwidth}
  \centering
  \includegraphics[width=\textwidth]{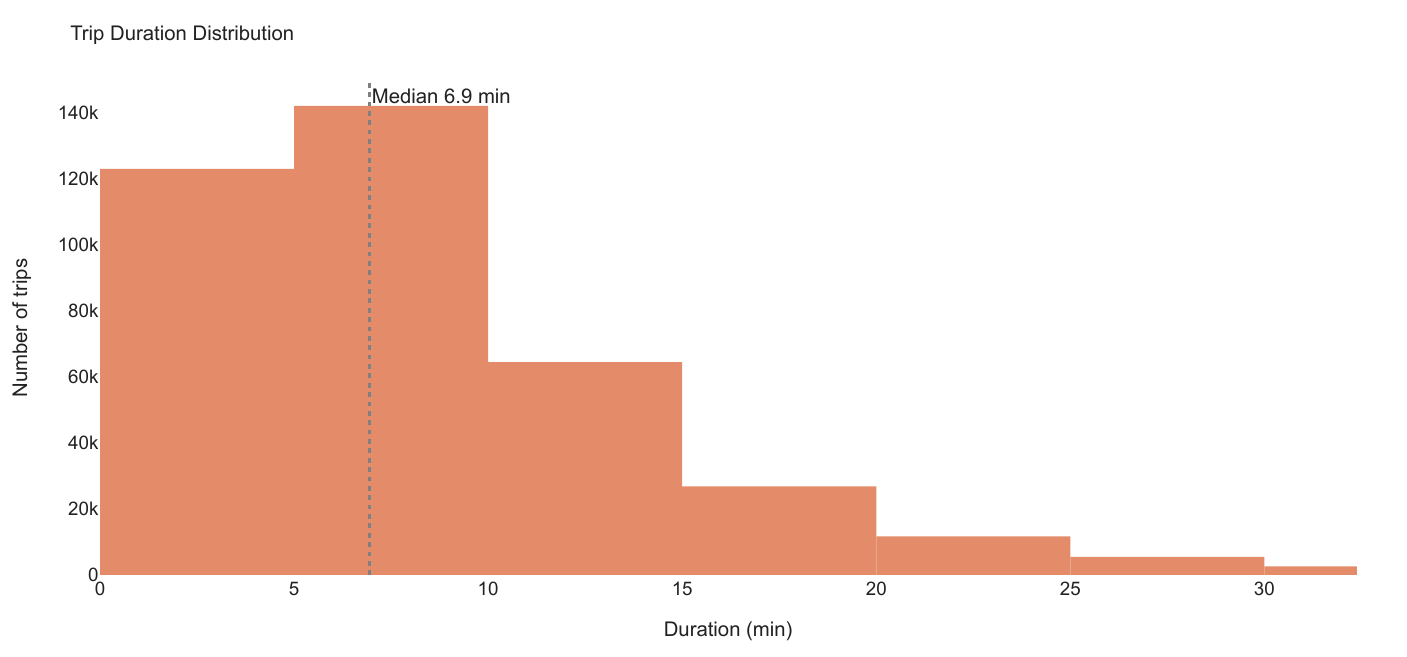}
  \caption{Trip duration distribution.}
  \label{fig:trip_dur}
\end{subfigure}
\caption{OD distance and trip duration distributions for Heilbronn (Apr~2025--Apr~2026).
Dashed vertical lines mark the respective medians (2.22~km and 6.9~min).}
\label{fig:od_duration}
\end{figure}

\begin{table}[htbp]
  \centering
  \caption{Start--end hotspot spatial overlap at increasing distance thresholds (Heilbronn, $n=200$).}
  \label{tab:overlap}
  \small
  \begin{tabular}{lcc}
  \hline
  \textbf{Distance threshold} & \textbf{Matched hotspots} & \textbf{Match rate} \\
  \hline
  $\leq$100\,m & 170 & 85.0\% \\
  $\leq$200\,m & 187 & 93.5\% \\
  $\leq$300\,m & 197 & 98.5\% \\
  $\leq$500\,m & 200 & 100.0\% \\
  \hline
  \multicolumn{3}{l}{\footnotesize Pearson $r = 0.972$ (start vs.\ end trip count), median separation 7\,m.}\\
  \hline
  \end{tabular}
  \end{table}
OD straight-line distances across the 1,585 hotspot pairs have a median of 2.22,km, with 88.3\% falling within 4,km. Trip durations show the same pattern: 87\% of trips complete within 15,min at a median of 6.9,min, implying a cruising speed of 16,km/h consistent with German e-scooter regulation. Table~\ref{tab:od_distance} reports the full breakdown by distance and duration band; Figure~\ref{fig:od_duration} visualises both distributions. Together they justify a 15-minute isochrone as the catchment area definition used throughout this framework.
  
\begin{table}[htbp]
\centering
\caption{OD straight-line distance and trip duration breakdown (Heilbronn).}
\label{tab:od_distance}
\small
\begin{tabular}{lrr|lrr}
\hline
\textbf{Distance band} & \textbf{OD pairs} & \textbf{Share}
  & \textbf{Duration band} & \textbf{Trips} & \textbf{Share} \\
\hline
$<$500\,m       &    17  &  1.1\% & $<$3\,min      &  44,042 & 11.6\% \\
500\,m--1\,km   &   207  & 13.1\% & 3--8\,min      & 175,792 & 46.4\% \\
1--2\,km        &   442  & 27.9\% & 8--15\,min     & 109,686 & 29.0\% \\
2--4\,km        &   732  & 46.2\% & 15--25\,min    &  38,494 & 10.2\% \\
$>$4\,km        &   187  & 11.8\% & $>$25\,min     &  10,843 &  2.9\% \\
\hline
\multicolumn{3}{l|}{\footnotesize Median 2.22\,km, mean 2.40\,km.}
  & \multicolumn{3}{l}{\footnotesize Median 6.9\,min, mean 8.7\,min.} \\
\hline
\end{tabular}
\end{table}

\begin{figure}[htbp]
\centering
\includegraphics[width=\textwidth, keepaspectratio]{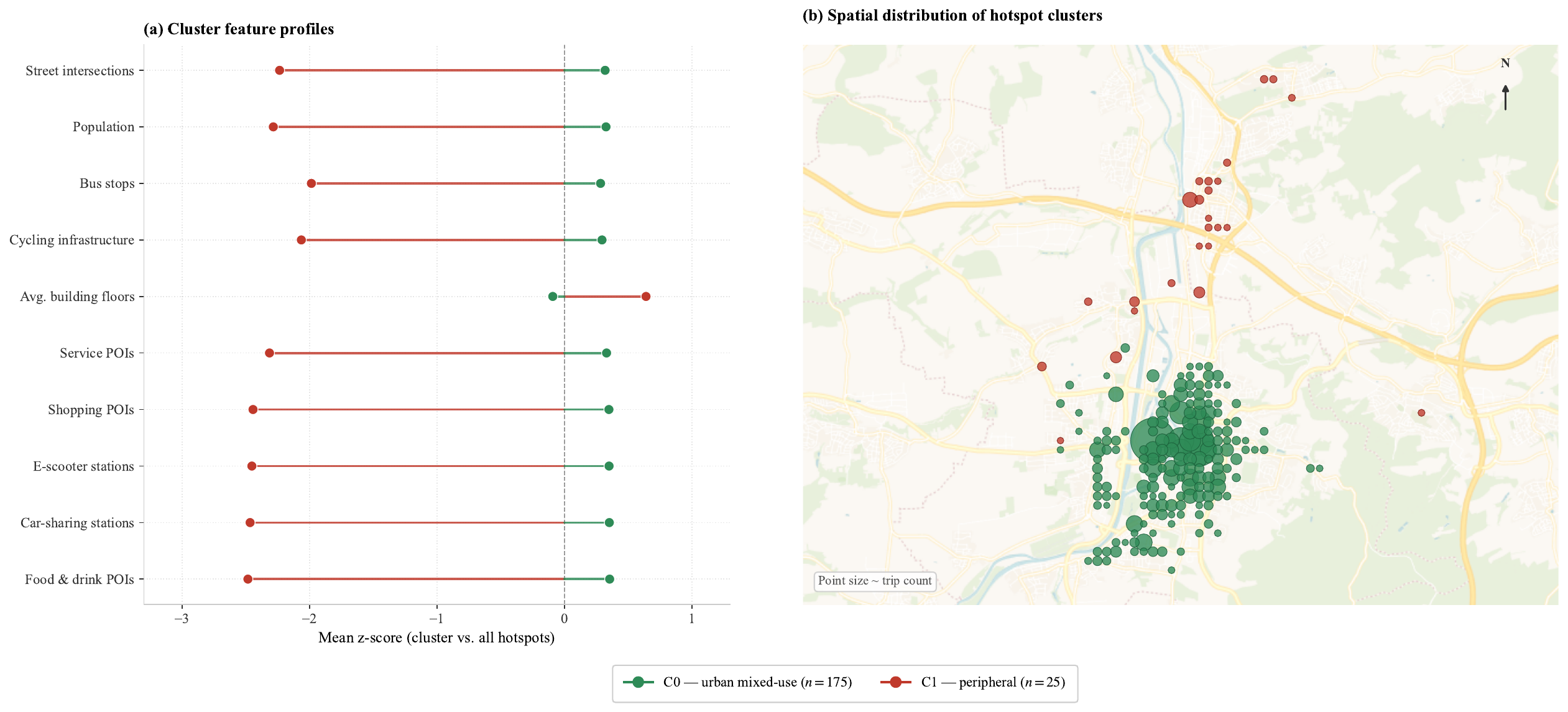}
\caption{Spatial clusters in Heilbronn ($k=2$): built-environment z-score profiles (left) and
geographic distribution (right). Point size proportional to trip count.}
\label{fig:spatial_clusters}
\end{figure}

KMeans partitioning selects $k=2$ (silhouette score). \textbf{C0} ($n=175$) scores above average on food \& drink POIs, car-sharing, e-scooter stations, and shopping POIs, consistent with a mixed-use urban core. \textbf{C1} ($n=25$) shows above-average building height but strongly negative scores across the same amenity categories, pointing to a peripheral, low-amenity context. Figure~\ref{fig:spatial_clusters} shows the z-score profiles and geographic distribution of both clusters.

\subsubsection{Phase 2: Agentic Causal Discovery}

\paragraph{Cluster Profiles}
The city--cluster profiles for Heilbronn's two spatial clusters are:

\begin{quote}
Heilbronn (130,093 inhabitants, 99.9\,km$^2$) is a mid-sized city in
northern Baden-W\"{u}rttemberg with predominantly flat terrain
(82\% flat roads) and three higher education institutions,
served by one e-scooter operator (Dott) over the study period.
Trip demand is heavily concentrated across the 200 hotspots
(Gini $G = 0.661$, Figure~\ref{fig:spatial_clusters}),
with \textbf{C0} forming the high-demand urban core
and \textbf{C1} the low-demand periphery.
\end{quote}

\paragraph{Statistical Pre-Analysis}
Table~\ref{tab:diagnostics} reports the statistical properties of
each cluster's feature matrix prior to causal discovery.

\textbf{C0} ($n = 175$, $p = 27$, $n/p = 6.48$):
the target variable \texttt{trip\_count} is strongly right-skewed
(skewness 6.2, kurtosis 50.1, Shapiro--Wilk $p \approx 0$),
requiring log-transformation before modelling.
Non-Gaussianity is widespread, with 23 of 27 features rejecting normality
at $\alpha = 0.05$ ($\rho_{\text{norm}} = 0.852$).
Multicollinearity is severe: 25 of 27 features have VIF $> 100$
and 48 feature pairs have Pearson $r > 0.80$
(maximum $r = 0.971$, \texttt{poi\_food\_drink} vs.\ \texttt{poi\_shopping}).

\textbf{C1} ($n = 25$, $p = 27$, $n/p = 0.93$):
the sample-to-feature ratio falls below one, making the feature matrix
rank-deficient. VIF scores are saturated for 25 of 27 features, and 125 feature pairs exceed $r > 0.80$.
\texttt{trip\_count} is non-Gaussian (skewness 2.7, kurtosis 7.65).

\begin{table}[htbp]
\centering
\caption{Data diagnostics for the two Heilbronn clusters.}
\label{tab:diagnostics}
\small
\setlength{\tabcolsep}{6pt}
\begin{tabular}{lcc}
\toprule
                              & \textbf{C0} & \textbf{C1} \\
\midrule
$n$ (hotspots)                & 175         & 25          \\
Features ($p$)                & 27          & 27          \\
$n/p$ ratio                   & 6.48        & 0.93        \\
Target skewness               & 6.20        & 2.74        \\
Target kurtosis               & 50.1        & 7.65        \\
$\rho_{\text{norm}}$          & 0.852       & 0.963       \\
High-VIF features             & 25/27       & 25/27       \\
High-corr pairs ($|r|>0.8$)  & 48          & 125         \\
\bottomrule
\end{tabular}
\end{table}

\paragraph{Orchestrator Configuration}
The orchestrator combines city-profile attributes and statistical
pre-analysis outputs to configure the causal discovery pipeline.

\noindent\textbf{Shared decisions:}
The skewed target (skewness $> 2$) requires log-transformation
of \texttt{trip\_count}. The 27-feature matrix is
decomposed into five semantic blocks (transportation $p=9$, POI $p=8$,
urban form $p=7$, population $p=3$, mobility supply $p=4$). City-profile-informed adjustments:
\texttt{pct\_flat\_terrain} is assigned low-signal status (82\% flat roads), \texttt{dormitory\_count} and \texttt{poi\_education} are retained in
the core group given Heilbronn's three registered universities.

\noindent\textbf{C0 (urban mixed-use):}
High non-Gaussianity ($\rho_{\text{norm}} = 0.852 $)
and sample size ($n = 175 \geq n_{\min}$) yield the
four-method set \{LiNGAM, DAGMA, NOTEARS, GRaSP\} with
consensus threshold $\geq 3/4$ across $B = 50$ bootstrap resamples.

\noindent\textbf{C1 (peripheral):}
The small sample ($n = 25 < n_{\min} = 50$) restricts the method set to
\{LiNGAM, DAGMA, NOTEARS\} with unanimity consensus ($= 3/3$).

\begin{figure}[htbp]
  \centering
  \includegraphics[width=\textwidth, keepaspectratio]{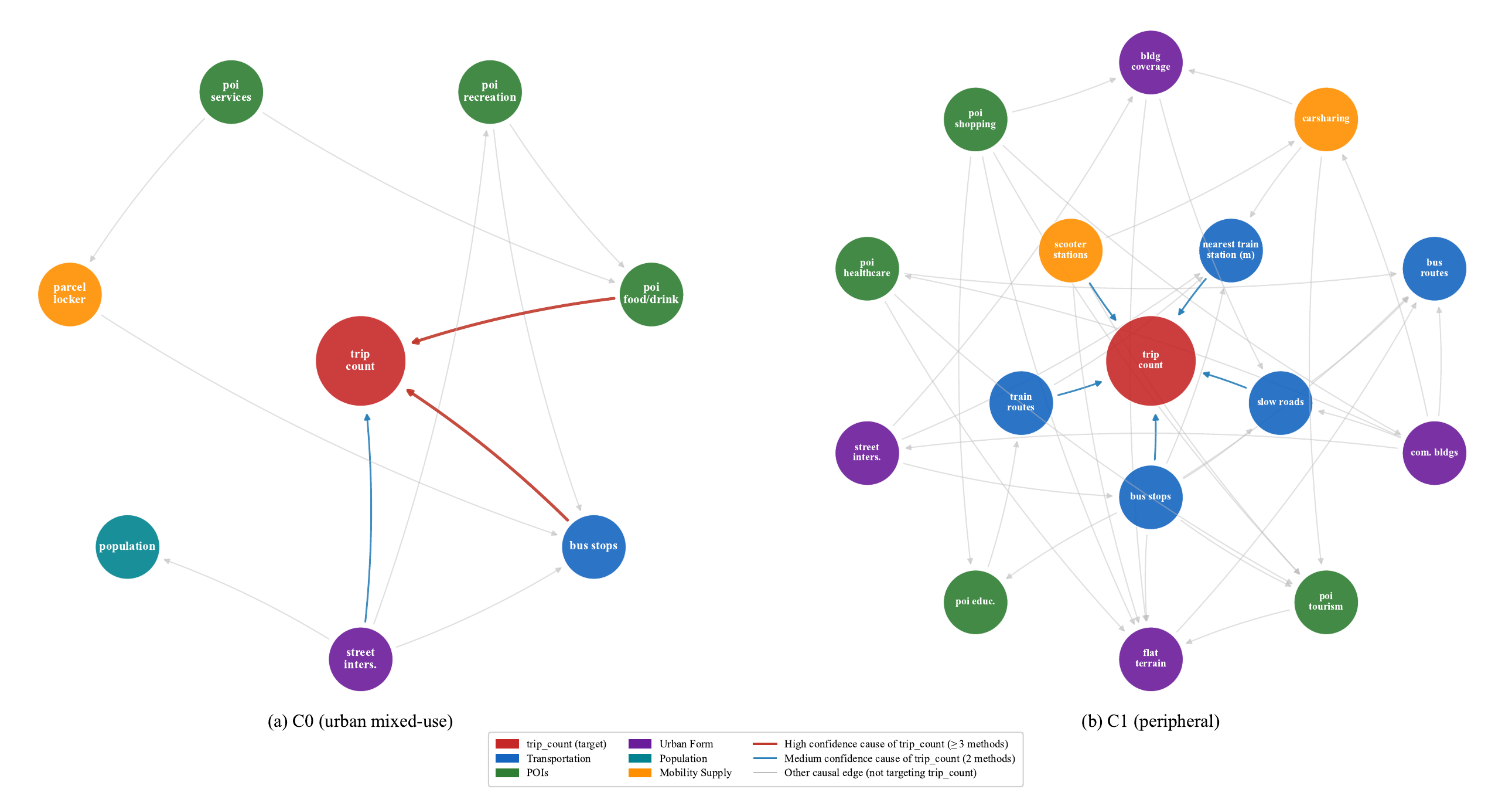}
  \caption{Consensus causal DAGs for C0 (urban mixed-use, left) and C1
  (peripheral, right), Heilbronn.}
  \label{fig:dags}
\end{figure}

\paragraph{Causal Discovery Results}
Figure~\ref{fig:dags} shows the consensus DAGs for both clusters.

\noindent\textbf{C0 (urban mixed-use):}
The four-method ensemble converged fully (bootstrap convergence $= 1.0$,
consensus rate $= 0.36$).
\texttt{street\_intersections} is the sole high-confidence cause, pointing to fine-grained urban grid structure
as the most robustly identified driver of trip intensity in the urban core.
Six medium-confidence direct causes were identified:
\texttt{bus\_stop\_count}, \texttt{poi\_food\_drink}, \texttt{poi\_services},
\texttt{population\_total}, \texttt{poi\_recreation}, and
\texttt{parcel\_locker\_count}. Taken together, trip generation in the urban core reflects the co-location
of transit access, daily service destinations, and pedestrian-scale street networks.

\noindent\textbf{C1 (peripheral):}
The three-method ensemble converged fully but at a lower consensus rate of 0.21, reflecting greater structural uncertainty at $n = 25$. No high-confidence cause was identified. Five medium-confidence drivers are: \texttt{bus\_stop\_count},
\texttt{train\_route\_count}, \texttt{nearest\_train\_station\_meter},
\texttt{pct\_slow\_roads}, and \texttt{scooter\_stations}.
The dominance of transit variables suggests peripheral hotspots function primarily as interchange nodes. \texttt{scooter\_\allowbreak stations} should be interpreted with caution, as operator placement may reflect observed demand rather than drive it.

\noindent\textbf{Cross-cluster comparison:}
C0 and C1 exhibit structurally different causal drivers: activity-driven in
the urban core (street connectivity, POI mix, population density) and
transit-driven in peripheral areas (rail and bus access).
This distinction has direct planning implications: hub reinforcement in C0
should prioritise street-level connectivity and mixed-use density, while
interventions in C1 should target first/last-mile integration at transit nodes.
The lower consensus rate in C1 and the absence of
high-confidence edges to the target reflect the inherent limitations of
small-sample causal discovery.

\subsubsection{Phase 3: Hub Siting and Planning Report}
\begin{figure}[htbp]
  \centering
  \includegraphics[width=\textwidth, height=0.45\textheight, keepaspectratio]{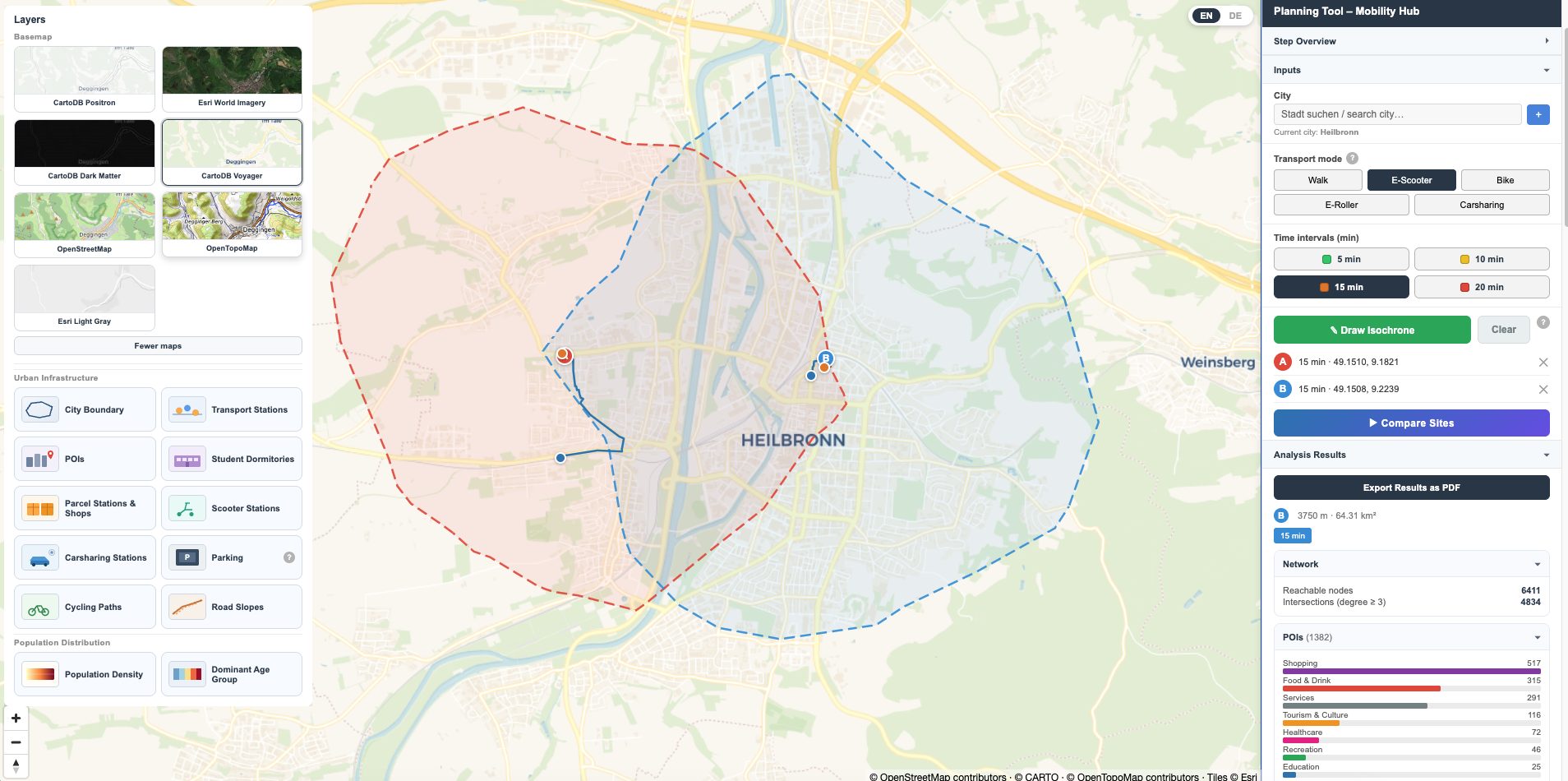}
  \caption{Planning tool: 15-minute e-scooter isochrones for Site~A (red) and Site~B (blue), Heilbronn.}
  \label{fig:planning_tool}
  \end{figure}

Two candidate sites illustrate the planning workflow (Figure~\ref{fig:planning_tool}): \textbf{Site~A} in the outer residential Saarland Street area and \textbf{Site~B} near two Heilbronn rail stations, both evaluated under a 15-minute e-scooter isochrone.
  
The planner selects the matching cluster template (urban mixed-use or peripheral), and the composite suitability score is computed by weighting the site's feature profile against the template's causal weight vector, combined with the age-weighted service demand from local population data.

\begin{figure}[htbp]
  \centering
  \includegraphics[width=\textwidth, keepaspectratio]{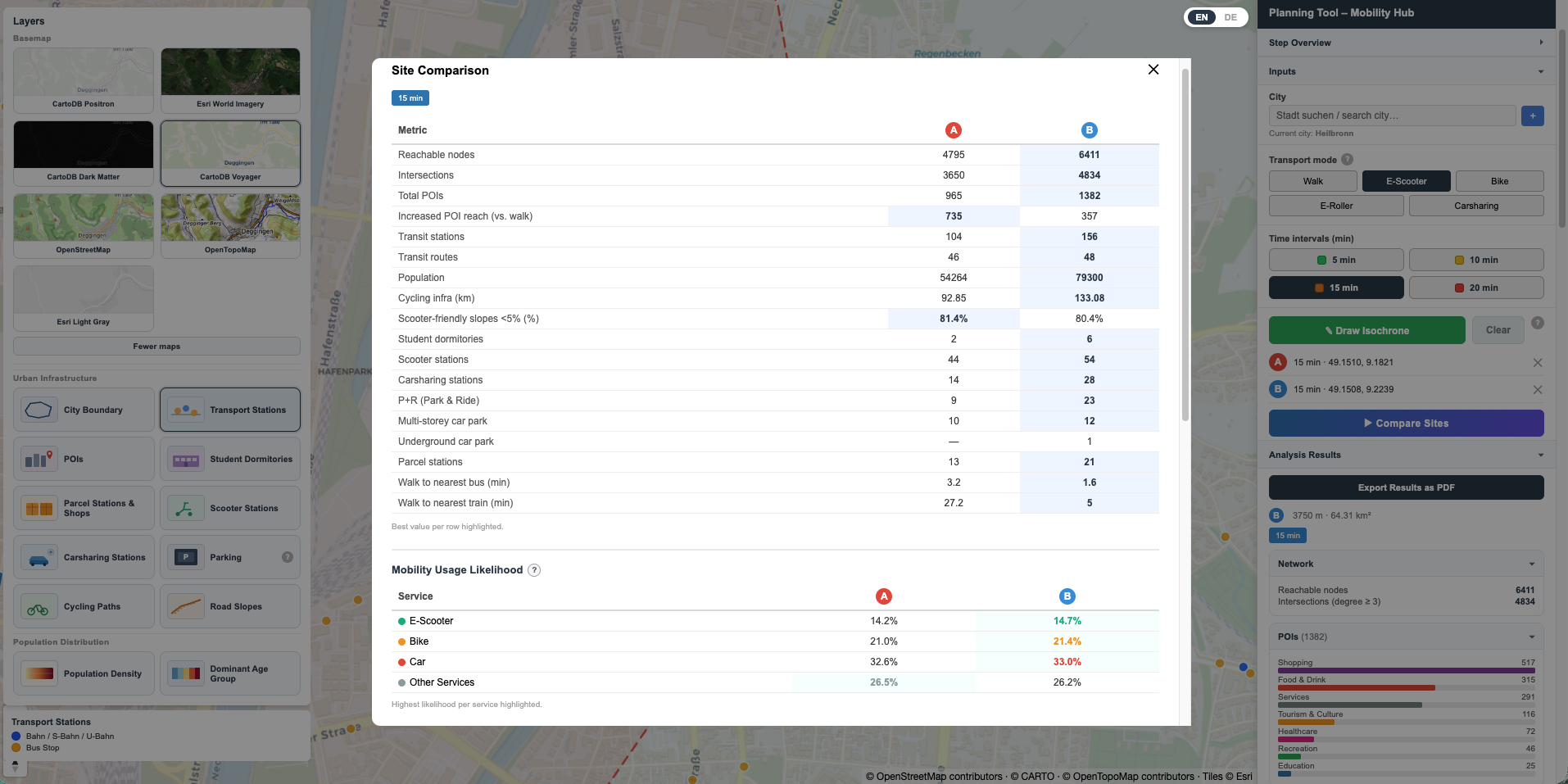}
  \caption{Site comparison panel: network metrics, POI counts, and mobility usage
  likelihood for Site~A and Site~B (15-minute e-scooter isochrone, Heilbronn).}
  \label{fig:site_comparison}
\end{figure}

The two sites differ markedly in their built-environment profiles
(Figure~\ref{fig:site_comparison}).
Site~B dominates on absolute metrics: larger network
(6{,}411 nodes vs.\ 4{,}795), greater population catchment (79{,}300 vs.\ 54{,}264),
higher total POI count (1{,}382 vs.\ 965), more cycling infrastructure (133.1~km vs.\ 92.9~km),
and better transit access (5~min walk to the nearest train station
versus 27.2~min for Site~A).
Site~A, however, achieves a higher incremental POI reach relative to walking
(735 additional POIs vs.\ 357 for Site~B), indicating that e-scooter mobility
unlocks greater accessibility gains at this peripheral location where walking
alternatives are weaker.
Age-calibrated mobility likelihood scores assign both sites similar predicted
e-scooter usage probabilities (14.7\% for Site~B vs.\ 14.2\% for Site~A). Companion Service Demand scores are also computed for each site. An LLM then synthesises all site profiles, causal weights, and service demand scores into a structured planning report.
  
 Beyond the demonstration case, two hub sites in Heilbronn were selected through the city's own planning process, informed by the tool's outputs: one peripheral (addressing last-mile connectivity, consistent with C1 drivers) and one in the urban core (consistent with C0 drivers). Both are currently under construction.

\subsection{Cross-City Scalability and City-Type Templates}
\label{sec5b}

This section extends the same pipeline across all 29 German cities
using GBFS open data. Two questions structure the analysis: which causal drivers
hold across diverse urban contexts, and how those drivers shift across city typologies.

\subsubsection{Consistent Causal Drivers Across 29 Cities}

\begin{figure}[htbp]
  \centering
  \includegraphics[width=\textwidth, height=0.6\textheight, keepaspectratio]{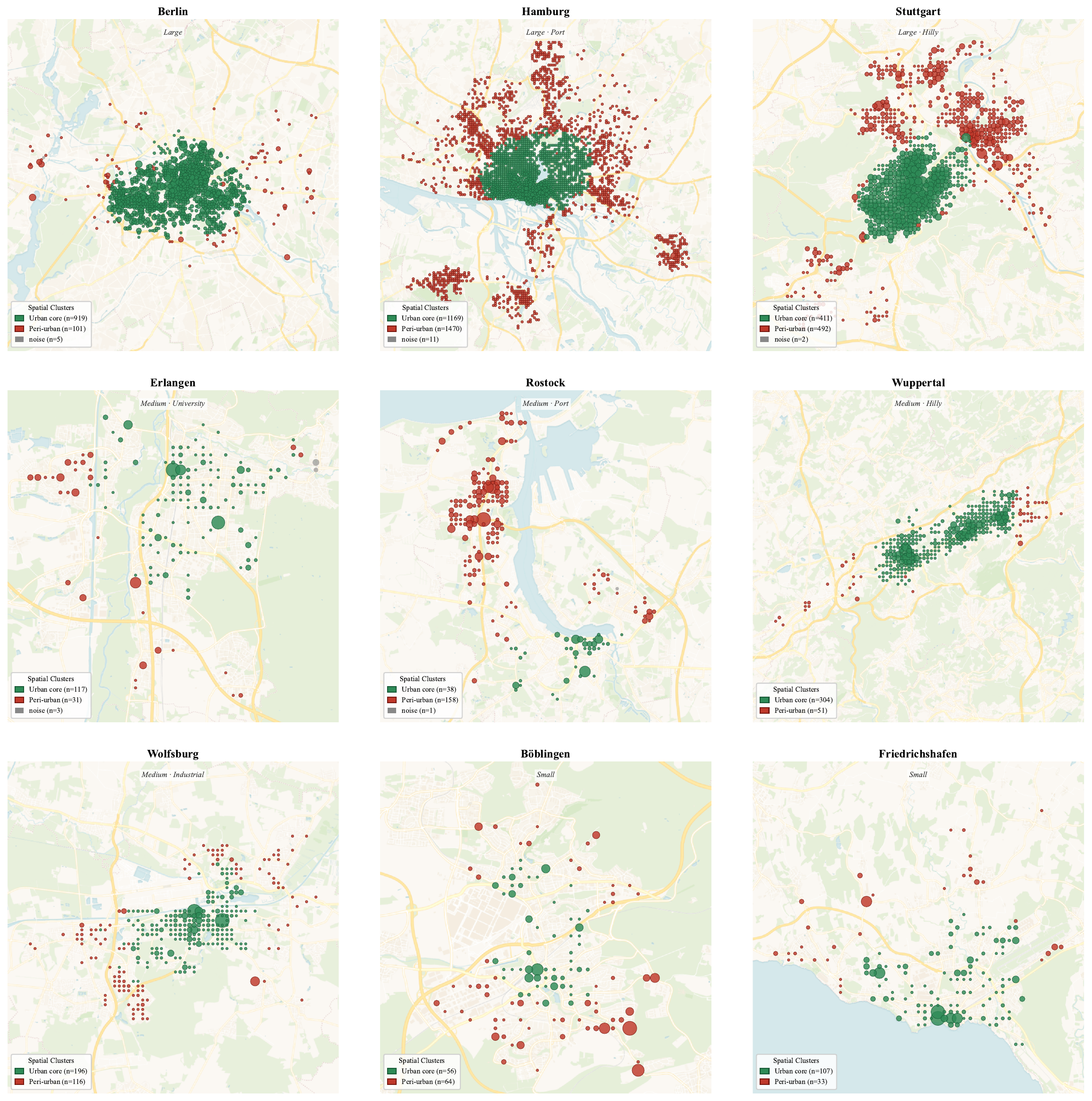}
  \caption{Spatial distribution of C0 (core high-activity) and C1 (peripheral
  low-activity) hotspot clusters for nine representative cities.
  Clusters are derived automatically by KMeans on environmental feature vectors.}
  \label{fig:nine_city_clusters}
\end{figure}

KMeans partitions hotspots into two clusters per city: C0 (core high-activity) and C1 (peripheral low-activity). All 29 cities yield a C0 cluster. Pforzheim's C1 is excluded due to insufficient hotspot count ($n = 7$), leaving 29 C0 and 28 C1 units (57 total). Figure~\ref{fig:nine_city_clusters} illustrates the spatial partitioning for nine representative cities.

Applied to all 57 units (11,916 hotspots), the pipeline yields a \textit{pass} verdict in 50 cases (87.7\%) and \textit{partial} in 7 (12.3\%), with no \textit{fail}. Full details of all 57 units are provided in \ref{app:all_units}. Table~\ref{tab:causal_drivers} reports the top-ten
drivers: recreation POIs rank first (40.4\%), followed by education POIs (36.8\%), three urban form features (building coverage, high-rise count, structured parking, each 33.3\%), street intersections (31.6\%), and train station count and proximity (both 29.8\%).   Occurrence ratios are moderate throughout, with no feature exceeding 50\%, reflecting the diversity of cities pooled. The city-type and cluster-type analyses below examine how drivers shift within more homogeneous subgroups.
  
\begin{table}[ht]
\centering
\caption{Ten most frequently identified direct causal drivers across 57 city--cluster
analysis units, grouped by feature block. Occurrence ratio is computed over all 57 units.}
\label{tab:causal_drivers}
\small
\begin{tabular}{llcc}
\hline
\textbf{Block} & \textbf{Feature} & \textbf{Units ($n$)} & \textbf{Occurrence ratio} \\
\hline
\multirow{3}{*}{POI}
  & Recreation POIs             & 23 & 40.4\% \\
  & Education POIs              & 21 & 36.8\% \\
  & Shopping POIs               & 17 & 29.8\% \\
\hline
\multirow{4}{*}{Urban form}
  & Building coverage (\%)      & 19 & 33.3\% \\
  & High-rise count             & 19 & 33.3\% \\
  & Structured parking count    & 19 & 33.3\% \\
  & Street intersections        & 18 & 31.6\% \\
\hline
\multirow{3}{*}{Transportation}
  & Train station count         & 17 & 29.8\% \\
  & Train station proximity (m) & 17 & 29.8\% \\
  & Bus stop count              & 16 & 28.1\% \\
\hline
\end{tabular}
\end{table}

C0 and C1 show systematically different driver profiles. In C0, recreation POIs and street intersections each appear in 45\% of units, followed by education POIs (41\%), train station count (38\%), and population total (34\%). In C1, built-form features dominate: building coverage, high-rise count, and structured parking each reach 32\%, while street intersections drop to 18\% and population total to 7\%. Car-sharing stations are more prominent in C1 (21\% vs.\ 10\%), reflecting intermodal last-mile access at the urban fringe. Full occurrence ratios for all 27 features are reported in \ref{app:cluster_freq}.
  
\subsubsection{City-Type Variation and Planning Templates}

The size and functional typologies in Table~\ref{tab:cities} produce
distinguishable causal signatures that translate into differentiated planning
recommendations.

\begin{figure}[htbp]
  \centering
  \includegraphics[width=\textwidth, height=0.55\textheight, keepaspectratio]{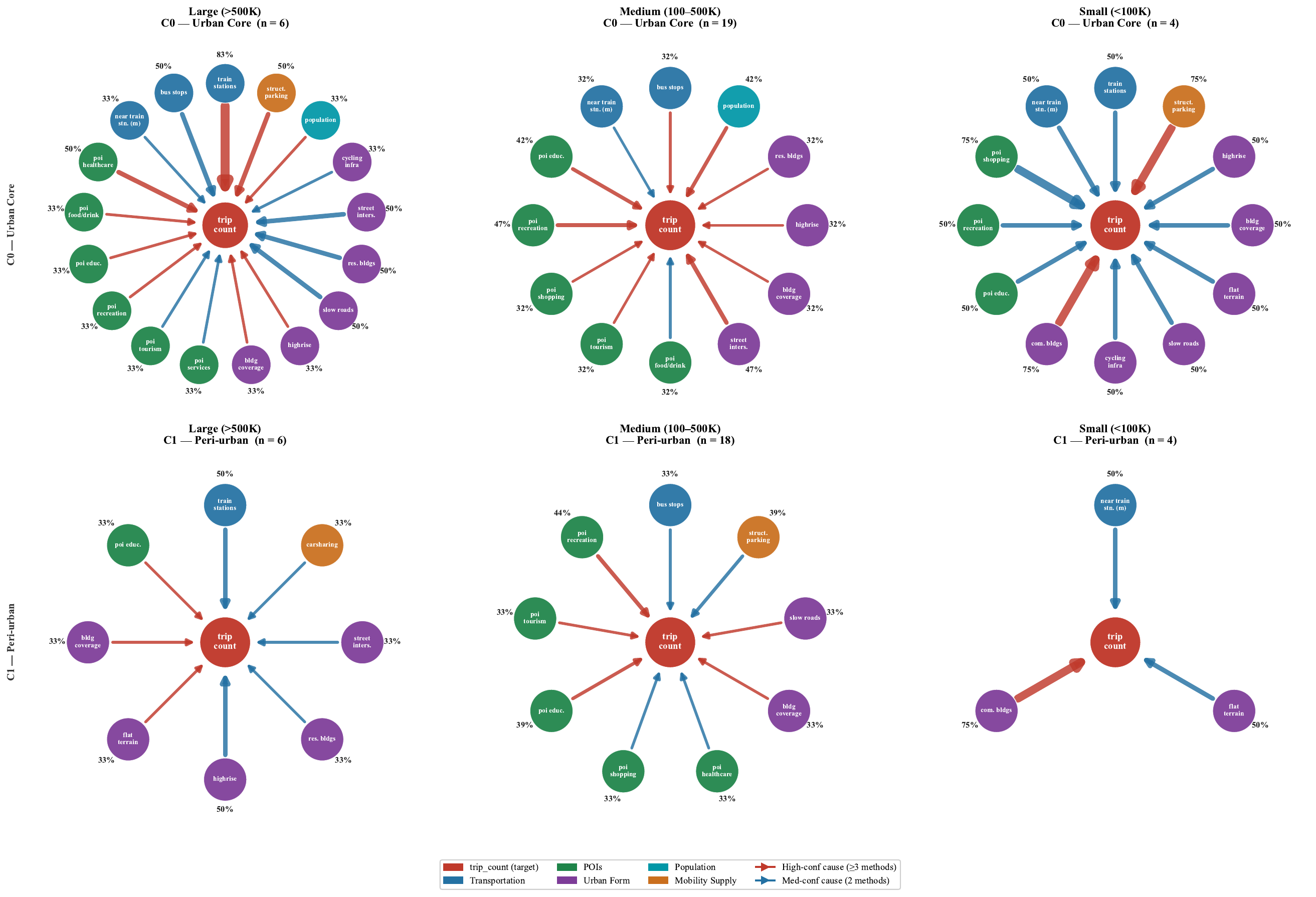}
  \caption{Consensus causal graphs grouped by city size class (columns:
  large $>$500K, medium 100--500K, small $<$100K) and cluster type (rows:
  C0 urban core, C1 peri-urban). Edge width reflects occurrence ratio across
  cities within each group. Red arrows: high-confidence causes ($\geq$3
  methods). Blue arrows: medium-confidence causes (2 methods).}
  \label{fig:dag_population_types}
\end{figure}

Figure~\ref{fig:dag_population_types} shows the consensus causal graphs
aggregated by city size and cluster type.

\textbf{Large cities} ($>$500K, $n = 6$): transit infrastructure dominates. Train station count
appears in 83\% of C0 units (the highest ratio of any feature in any size class), with bus stops,
train station proximity, street intersections, residential buildings, and slow road share each at
50\%. In C1, high-rise count and train stations lead (50\% each), alongside building coverage, flat
terrain, residential buildings, street intersections, and car-sharing stations (each 33\%). The large-city causal structure reflects major rail interchanges as primary
demand anchors in core areas, while peripheral demand retains a transit and
density signal alongside intermodal access.

\textbf{Medium cities} (100K--500K, $n = 19$) show the most distributed driver set, with no feature exceeding 50\% occurrence in either cluster type. In C0, recreation POIs and street intersections rank joint first (47\%), followed by education POIs and train station proximity.
In C1, recreation POIs lead at 44\%, with education POIs and structured parking at 39\%, and bus stops, tourism, shopping, healthcare, building coverage, and slow roads each at 33\%. This is a direct consequence of the group's composition: the 19 medium cities
span regional administrative centres, university towns, port cities, and post-industrial cities, each with a different dominant driver. The diversity of the group suppresses any single signal at the aggregate level,
and city-specific output should be consulted for planning in this size class.

\textbf{Small cities} ($<$100K, $n = 4$): the most concentrated causal graphs. In C0, train
station count, train station proximity, structured parking, and commercial buildings each appear
in 75\% of units with high confidence. In C1, commercial buildings dominate (75\%), with train
station proximity and flat terrain at 50\%. Hub positioning relative to the main rail station and
adjacent commercial activity is the primary planning decision for this size class.
  
\begin{figure}[htbp]
  \centering
  \includegraphics[width=\textwidth, height=0.55\textheight, keepaspectratio]{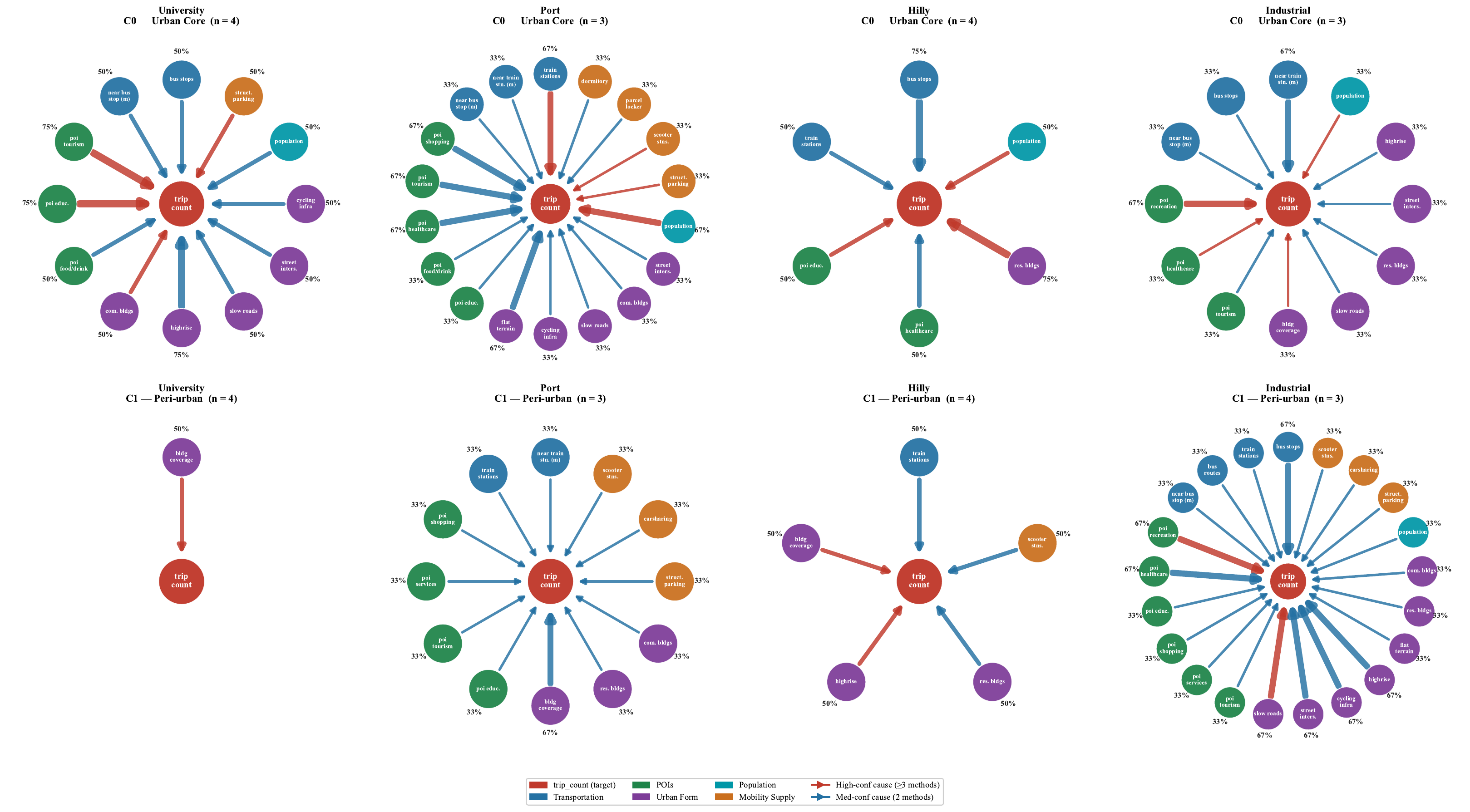}
  \caption{Consensus causal graphs grouped by functional city type (columns:
  university, port, hilly, industrial) and cluster type (rows: C0 urban core,
  C1 peri-urban). Edge width reflects occurrence ratio across cities within
  each group. Red arrows: high-confidence causes ($\geq$3 methods).
  Blue arrows: medium-confidence causes (2 methods).}
  \label{fig:dag_functional_types}
\end{figure}

Figure~\ref{fig:dag_functional_types} shows the consensus causal graphs
aggregated by functional city type and cluster type.

\textbf{University cities}: education POIs, tourism POIs, and commercial buildings are high-confidence C0 drivers in 75\% of units; cycling infrastructure appears in 50\%. The tourism POI signal reflects cultural and leisure amenities rather than visitor tourism. In C1, only building coverage appears (50\%), consistent with university-specific drivers being spatially concentrated in the urban core.

\textbf{Port cities}: the most diverse C0 driver set, with train stations, population, shopping,
tourism, and healthcare POIs, and flat terrain each appearing in 67\% of units. The flat terrain
signal reflects the coastal topography shared by all three cities. In C1, building coverage dominates (67\%), with remaining features scattered at 33\%.

\textbf{Hilly cities} show a
sparse C0 causal structure. Bus stops and residential buildings each appear in
75\%, population, train stations, education
POIs, and healthcare POIs each in 50\%.
Flat terrain does not emerge as a consistent group-level driver in C0, suggesting
that terrain accessibility effects are city-specific rather than uniform across
hilly cities.
In C1, building coverage, high-rise count, residential buildings, train stations,
and scooter stations each appear in 50\% of units.

\textbf{Industrial cities}: recreation POIs and train station proximity each appear in 67\% of C0
units, suggesting demand concentrates in leisure and mixed-use zones rather than employment areas.
C1 shows the most complex driver set of any functional type: bus stops, recreation POIs,
healthcare POIs, slow road share, street intersections, cycling infrastructure, and high-rise
count each at 67\%, reflecting a combination of transit access, walkability, and built-environment density typical of post-industrial peripheries.

\FloatBarrier
\section{Discussion}
\label{sec6}

\subsection{Recommendations and Implications}

\textbf{For hub siting.}
The causal weight library should be treated as a structured prior rather than a
deterministic ranking: modifiable drivers are intervention
targets, while non-modifiable signals such as building coverage serve as site filters.
Planners can apply the matched city-type template directly and reserve detailed
city-specific analysis for sites that deviate substantially from template predictions.

\textbf{For facility specification.}
A single generic facility list will systematically underserve part of the user base.
Cities with younger user concentrations around universities or nightlife areas should
weight transit connectivity and charging access more heavily in their hub design briefs.
Cities with older or more diverse populations should prioritise covered waiting areas,
seating, and physical accessibility.
The demographic calibration step in Phase~3 is designed to make this adjustment
explicit and reproducible at each candidate site.

\textbf{For multi-modal planning.}
The citizen preference survey reveals a clear asymmetry: public transport integration is rated the
most important hub feature nationally (78\%), while e-scooters rank lowest (25--28\%). This study
focuses on e-scooters because trip data are available at scale for this mode, but the pipeline is
mode-agnostic: the same hotspot detection, feature extraction, and causal discovery steps apply to
car-sharing, bike-sharing, or on-demand transit given equivalent spatial data. Extending the
analysis as open data become available would yield a more complete causal account of multi-modal
hub demand.
  
\textbf{For data and policy.}
GBFS feed publication is sufficient for planning-grade causal analysis without bilateral data-sharing agreements. Transport authorities should consider requiring high-frequency GBFS feeds as a standard licence condition. Standardising snapshot frequency and vehicle identifier consistency across operators would further reduce trip inference uncertainty.

\subsection{Limitations}

\textbf{LLM reproducibility.}
The analytical steps within each run are deterministic, but orchestrator decisions
on borderline cases depend on LLM inference and may differ between invocations.
All decisions were logged and random seeds fixed where applicable,
but full reproducibility of LLM-mediated analysis at this scale remains an open challenge.

\textbf{Partial consensus.}
The 12.3\% partial rate reflects residual collinearity in urban feature matrices
that bootstrap consensus does not fully resolve.
Edge directions in partial-quality graphs carry uncertainty and should be
treated as indicative rather than confirmed causes.

\textbf{Single-cluster exclusion.}
Pforzheim's C1 cluster was excluded due to insufficient hotspot count ($n = 7$).
The exclusion is systematic and reproducible but reduces spatial coverage
for that city.

\textbf{Temporal scope.}
The cross-city analysis uses approximately 40 days of GBFS data per city.
Seasonal variation may affect causal estimates, and a single observation window
is sensitive to atypical weather or event-driven demand.
Longer GBFS collection windows would improve temporal robustness.

\textbf{Geographic scope.}
The framework is validated exclusively within the German urban context,
where regulatory conditions, street design standards, and GBFS adoption
are relatively homogeneous.
Transferability to other regulatory environments or urban morphologies
remains to be tested.

\subsection{Expert Validation}
\label{sec5c}

The framework has been deployed as a web-based planning tool open to all German
municipalities. Prior to public release, 22 city planners spanning all city size
classes and functional typologies in the study sample have registered to compare
the tool's recommendations against their local planning knowledge. Evaluation scores
and qualitative feedback will be reported in a follow-up publication.

\FloatBarrier
\section{Conclusion}
\label{sec9}

This paper presented a three-phase agentic framework for e-scooter mobility hub planning
across 29 German cities, addressing four gaps in the literature: limited trip data
availability, reliance on correlational rather than causal evidence, the absence of
demographic calibration for facility design, and the gap between analytical outputs
and actionable planning decisions.

On data access, GBFS-inferred hotspots matched 95.5\% of the operator's MDS-recorded
hotspots within 300\,m ($r = 0.928$) in Heilbronn, establishing that open GBFS data
can replace proprietary trip records for hotspot detection.
Across all 57 city-cluster units, causal discovery yielded a pass verdict in 87.7\%
of cases. The 7 partial verdicts reflect collinearity in the feature matrices, not
data insufficiency.
The full pipeline is therefore applicable to any city with a public GBFS feed.

On causal evidence, recreation POIs (40.4\%), education POIs (36.8\%), and building
coverage (33.3\%) are the most consistently identified direct drivers across the full
dataset.
Stratification by city type and cluster reveals systematic variation beneath these
aggregates: core hotspot demand is shaped by activity access and transit proximity,
while peripheral demand responds to built form and density.
For large and small cities this variation is regular enough to support transferable
causal planning templates. Medium cities require city-specific analysis due to their
internal compositional diversity.

On facility design, the citizen preference survey confirms that age composition
materially affects which hub facilities are warranted.
Younger users prioritise transit connectivity and charging access, while older users
weight covered waiting areas and physical accessibility.
The demographic calibration step translates these differences into site-specific
infrastructure recommendations derived from the actual catchment population.

On implementation, both sites recommended by the framework in Heilbronn were approved
and are currently under construction, completing the full pathway from open data to
building permit.
The causal weight library has been deployed as a planning tool open to all German
municipalities, with a structured evaluation by 22 registered city planners under way.

Three directions remain for further work. Post-construction monitoring of the Heilbronn
hubs will test causal predictions against observed ridership. Extension to shared e-bikes
and car-sharing will determine how much of the causal template transfers across modes.
Application outside the German context will test generalisability to different regulatory
environments and urban morphologies.
\section*{CRediT authorship contribution statement}

\textbf{Meng Jin:} Writing -- original draft, Visualization, Software, Methodology, Investigation, Formal analysis.
\textbf{Melanie Handrich:} Writing -- review \& editing, Funding acquisition, Project administration, Conceptualization.
\textbf{Simone Martinenz:} Writing -- review \& editing, Funding acquisition, Project administration.
\textbf{Nicholas Hoeser:} Writing -- review \& editing, Data curation, Formal analysis.
\textbf{Ziyue Li:} Writing -- review \& editing, Conceptualization, Methodology, Supervision.

\section*{Data availability}

The data used in this study will be made available on request.
The planning tool dashboard will be publicly accessible to all users from June 2026.
The source code underlying the planning tool will be released as open source in December 2026.

\section*{Declaration of competing interest}

The authors declare that they have no known competing financial interests or personal relationships that could have appeared to influence the work reported in this paper.

\section*{Acknowledgments}

This work was conducted within the Scoop2City project (``Data-Driven Service Ecosystem for the City-Compatible Establishment and Operation of Shared Mobility Offerings for Municipalities''), carried out by the Fraunhofer Institute for Industrial Engineering IAO and funded by the German Federal Ministry for Digital Affairs and Transport (BMDV) through the mFUND innovation programme (Grant No.: 19F2261B). The authors thank the project partners and the City of Heilbronn for their support and collaboration.

\appendix
\setlength{\floatsep}{4pt plus 1pt minus 1pt}
\setlength{\textfloatsep}{6pt plus 1pt minus 2pt}
\setlength{\intextsep}{4pt plus 1pt minus 1pt}
\setlength{\abovecaptionskip}{4pt}
\setlength{\belowcaptionskip}{2pt}
\section{Sensitivity Analyses}
\label{app:sensitivity}

\subsection{Hotspot Detection Parameters ($\varepsilon$ and $\tau$)}
\label{app:sensitivity:hotspot}

\begin{figure}[!ht]
\centering
\includegraphics[width=\linewidth]{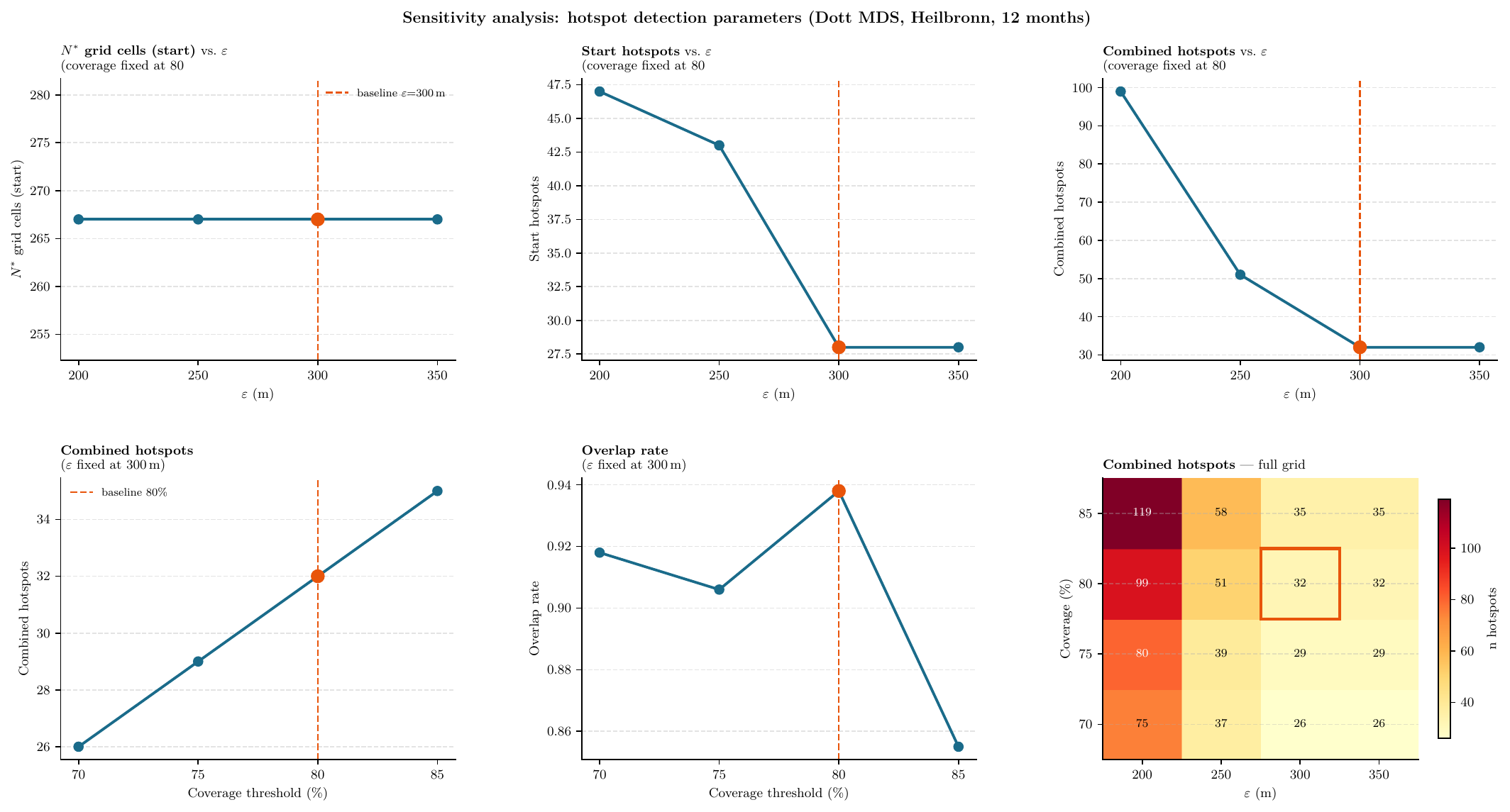}
\caption{Sensitivity of hotspot detection metrics to $\varepsilon$
  (top row, coverage fixed at 80\%) and to the coverage threshold
  (bottom row, $\varepsilon$ fixed at 300\,m).
  The heatmap (bottom right) shows combined hotspot counts across
  the full parameter grid. The orange box marks the chosen baseline
  $(\varepsilon=300\,\text{m},\;\tau=80\%)$.}
\label{fig:sensitivity}
\end{figure}

We vary $\varepsilon \in \{200, 250, 300, 350\}$\,m and $\tau \in \{70\%, 75\%, 80\%, 85\%\}$
independently on the Heilbronn MDS dataset, holding the other parameter at its baseline.
Full results are in Table~\ref{tab:sensitivity} and Figure~\ref{fig:sensitivity}.

$\varepsilon = 300$\,m is the minimum radius at which hotspot count and
start--end overlap converge: results at 350\,m are identical, confirming convergence.
$\tau = 0.80$ maximises overlap (93.8\%) at the inflection of the coverage curve;
raising $\tau$ to 85\% adds spatially diffuse, low-frequency cells that reduce overlap to 85.5\%.

\begin{table}[!ht]
\centering
\caption{Sensitivity analysis results across all $\varepsilon$--$\tau$
  combinations. The baseline is highlighted in bold.
  $N^*$: grid cells selected (start). HS: combined hotspots after DBSCAN.
  OL: symmetric start--end overlap rate. T10: fraction of trips in
  top-10 combined hotspots.}
\label{tab:sensitivity}
\small
\setlength{\tabcolsep}{6pt}
\begin{tabular}{ccrrrrrr}
\toprule
$\varepsilon$ (m) & $\tau$ & $N^*$ & HS (start) & HS (end) & HS (combined) & OL & T10 \\
\midrule
200 & 70\% & 178 & 39 & 79 & 75  & 72.7\% & 50.3\% \\
200 & 75\% & 218 & 40 & 92 & 80  & 77.7\% & 54.0\% \\
200 & 80\% & 267 & 47 & 108& 99  & 74.9\% & 56.3\% \\
200 & 85\% & 329 & 57 & 128& 119 & 72.3\% & 58.5\% \\
\midrule
250 & 70\% & 178 & 33 & 32 & 37  & 86.2\% & 64.8\% \\
250 & 75\% & 218 & 36 & 38 & 39  & 94.7\% & 69.7\% \\
250 & 80\% & 267 & 43 & 48 & 51  & 88.2\% & 73.4\% \\
250 & 85\% & 329 & 52 & 53 & 58  & 89.5\% & 78.7\% \\
\midrule
\textbf{300} & \textbf{70\%} & \textbf{178} & \textbf{23} & \textbf{25} & \textbf{26} & \textbf{91.8\%} & \textbf{67.2\%} \\
\textbf{300} & \textbf{75\%} & \textbf{218} & \textbf{27} & \textbf{26} & \textbf{29} & \textbf{90.6\%} & \textbf{71.7\%} \\
\textbf{300} & \textbf{80\%} & \textbf{267} & \textbf{28} & \textbf{32} & \textbf{32} & \textbf{93.8\%} & \textbf{76.7\%} \\
300 & 85\% & 329 & 32 & 29 & 35  & 85.5\% & 82.4\% \\
\midrule
350 & 70\% & 178 & 23 & 25 & 26  & 91.8\% & 67.2\% \\
350 & 75\% & 218 & 27 & 26 & 29  & 90.6\% & 71.7\% \\
350 & 80\% & 267 & 28 & 32 & 32  & 93.8\% & 76.7\% \\
350 & 85\% & 329 & 32 & 29 & 35  & 85.5\% & 82.4\% \\
\bottomrule
\end{tabular}
\end{table}

Figure~\ref{fig:sensitivity} visualises the same results as line plots and a heatmap.

\subsection{Minimum Cluster Size Threshold ($n_{\min} = 50$)}
\label{app:n_threshold}

We validate $n_{\min} = 50$ via a subsampling experiment on Heilbronn C0
($n_{\text{full}} = 164$). Because the global feature set is severely
multicollinear (VIF $> 10$ for 26 of 27 features), GRaSP is fitted
independently on each of the five semantic blocks defined in
Table~\ref{tab:features} (Section~\ref{sec3}).
For each $n \in \{20, 30, 40, 50, 60, 75, 100\}$, $K = 20$
random subsamples and $B = 50$ bootstrap resamples yield mean block
stability $\bar{s}$ and target-directed stability $\bar{s}^{\text{tgt}}$
(edges into \texttt{trip\_count}).

Full results are in Table~\ref{tab:n_sensitivity}.

\begin{table}[!ht]
\centering
\caption{Bootstrap stability as a function of subsample size $n$.
  $\bar{s}$: mean overall stability (across 5 blocks and 20 subsamples).
  $\sigma_s$: standard deviation across 20 subsamples.
  $\bar{s}^{\text{tgt}}$: mean stability of edges directed into
  \texttt{trip\_count} (n.a. = no such edge found in any reference graph).
  Per-block columns show the block-level mean.
  Baseline threshold ($n = 50$) in bold.}
\label{tab:n_sensitivity}
\small
\setlength{\tabcolsep}{5pt}
\begin{tabular}{r rr r rrrrr}
\toprule
$n$ & $\bar{s}$ & $\sigma_s$ & $\bar{s}^{\text{tgt}}$
    & Trans. & POI & Urban & Pop. & Mobi. \\
\midrule
 20 & 0.107 & 0.062 & 0.12$^\dagger$ & 0.106 & 0.093 & 0.316 & 0.000 & 0.022 \\
 30 & 0.111 & 0.063 & n.a.           & 0.100 & 0.143 & 0.313 & 0.000 & 0.000 \\
 40 & 0.138 & 0.053 & n.a.           & 0.174 & 0.102 & 0.365 & 0.000 & 0.050 \\
\textbf{50} & \textbf{0.143} & \textbf{0.066} & \textbf{0.26}
    & 0.199 & 0.186 & 0.285 & 0.026 & 0.018 \\
 60 & 0.175 & 0.068 & 0.29 & 0.213 & 0.289 & 0.332 & 0.042 & 0.000 \\
 75 & 0.201 & 0.067 & 0.35 & 0.302 & 0.248 & 0.367 & 0.088 & 0.000 \\
100 & 0.224 & 0.048 & 0.39 & 0.374 & 0.218 & 0.452 & 0.078 & 0.000 \\
\bottomrule
\multicolumn{9}{l}{%
  $^\dagger$ Observed in only 1 of 20 subsamples, treated as unreliable.
  Trans.\ = transportation. Mobi.\ = mobility supply.}
\end{tabular}
\end{table}

At $n < 50$, GRaSP produces no demand-directed edges in any subsample or block.
At $n = 50$, $\bar{s}^{\text{tgt}} = 0.26$ first appears (transportation and
population blocks), rising to 0.39 at $n = 100$. Overall stability increases
from 0.107 ($n = 20$) to 0.224 ($n = 100$), accelerating more steeply above
the threshold. 

\subsection{Sensitivity to the Normality Failure Rate Threshold ($\rho_{\text{norm}} = 0.5$)}
\label{app:normality_threshold}

GRaSP is added to the pipeline when $n \geq 50$ and the normality failure rate
$\hat{\rho}$ (fraction of features rejecting Shapiro--Wilk normality at $p < 0.05$)
exceeds 0.5. We justify this choice on three grounds.

\begin{figure}[!ht]
  \centering
  \includegraphics[width=\textwidth]{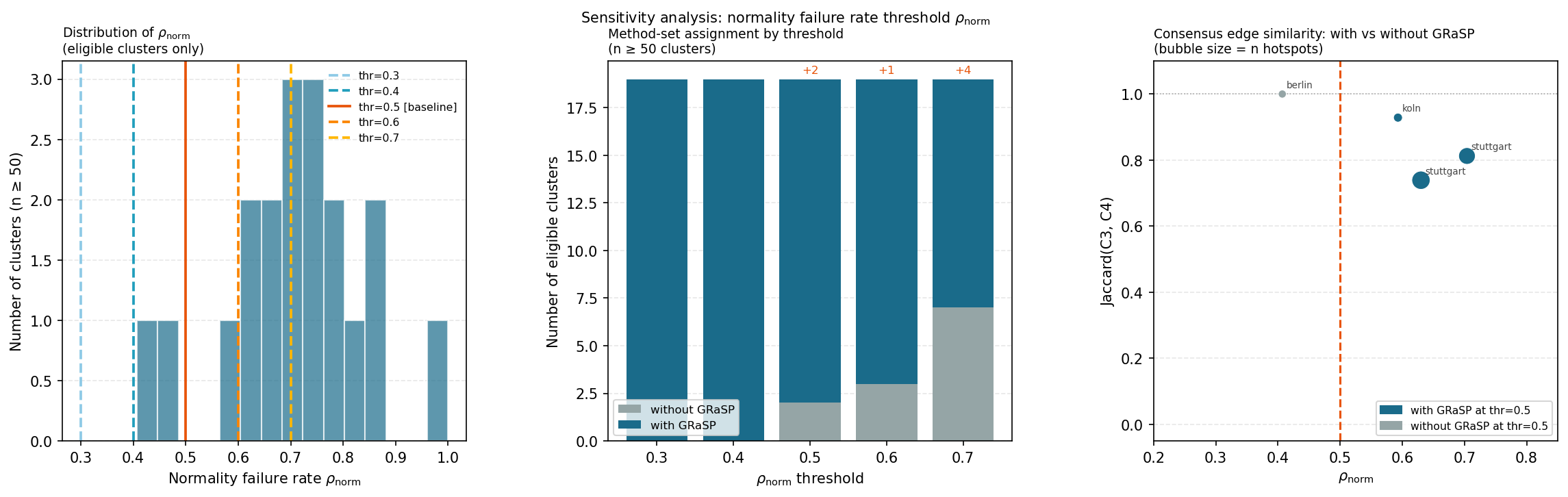}
  \caption{Sensitivity to $\rho_{\text{norm}}$ threshold (baseline 0.5, dashed orange).
    \emph{Left}: $\hat{\rho}$ distribution across eligible clusters ($n \geq 50$).
    \emph{Centre}: fraction assigned with GRaSP at each threshold.
    \emph{Right}: Jaccard similarity with vs without GRaSP as a function of $\hat{\rho}$
    (bubble size $\propto n$).}
  \label{fig:normality_threshold}
\end{figure}

\textbf{Data-driven natural break.}
Among the 19 eligible clusters ($n \geq 50$), the largest gap in observed $\hat{\rho}$
values falls between K\"{o}ln C0 ($\hat{\rho} = 0.481$) and K\"{o}ln C1
($\hat{\rho} = 0.593$), a gap of approximately 0.11.
The threshold 0.5 lies within this interval
(Figure~\ref{fig:normality_threshold}, left), so small perturbations to the threshold
do not reassign any cluster.

\textbf{Semantic clarity.}
$\hat{\rho} > 0.5$ means that a majority of features significantly depart from
normality, providing an intuitive rationale for introducing a non-parametric algorithm.

\textbf{Result robustness.}
Moving the threshold from 0.5 to 0.6 reassigns only one cluster
(K\"{o}ln C1, $\hat{\rho} = 0.593$), and the Jaccard similarity between its
consensus edge sets with and without GRaSP is 0.929
(Figure~\ref{fig:normality_threshold}, right), indicating that the boundary
assignment has negligible effect on the final causal structure.

Of the 19 eligible units, 17 include GRaSP at the baseline threshold.
For four clusters spanning $\hat{\rho} \in [0.3, 0.7]$, Table~\ref{tab:normality_comparison}
reports the full comparison. Jaccard similarity ranges from 0.74 to 1.00 and
$\Delta_{\text{tgt}} = 0$ for three of four clusters. In Stuttgart C0
($\hat{\rho} = 0.70$), the pipeline without GRaSP identifies one additional target driver
(\texttt{population\_total} $\to$ \texttt{trip\_count}) absent from the four-method
pipeline, yielding $\Delta_{\text{tgt}} = -1$.

\begin{table}[!ht]
\centering
\caption{Consensus edge sets with vs without GRaSP for clusters in the sensitive zone
  ($\hat{\rho} \in [0.3, 0.7]$, $n \geq 50$).
  $J$: Jaccard similarity. $\Delta_{\text{tgt}}$: change in edges into \texttt{trip\_count}.}
\label{tab:normality_comparison}
\small
\begin{tabular}{lrrrrrc}
\toprule
City-cluster & $\rho_{\text{norm}}$ & $n$ & $|E_{\text{no}}|$ & $|E_{+}|$ & $J$ & $\Delta_{\text{tgt}}$ \\
\midrule
Berlin C0    & 0.41 &  95 & 17 & 17 & 1.00 & 0 \\
Cologne C1   & 0.59 & 118 & 13 & 14 & 0.93 & 0 \\
Stuttgart C1 & 0.63 & 492 & 20 & 20 & 0.74 & 0 \\
Stuttgart C0 & 0.70 & 411 & 15 & 14 & 0.81 & $-1$ \\
\bottomrule
\end{tabular}
\end{table}

\subsection{Sensitivity to the $n/p$ Exclusion Threshold ($\tau = 5$)}
\label{app:np_threshold}

Constraint-based methods (PC, FCI, GES) are excluded when $n/p < \tau$.
We re-ran the full block-level analysis on B{\"o}blingen, Heilbronn, and Hamburg
under $\tau \in \{3, 5, 7, \infty\}$.
High-confidence causal features are identical across $\tau = 3$, 5, and 7:
within-block $n/p$ ratios either exceed 7 (large/medium cities) or fall below 5
(specific small-city blocks), leaving no case where the eligible method set
differs between these values.
Only $\tau = \infty$ diverges: in Hamburg C0, permanently excluding PC and FCI
reduces identified causes from \{\texttt{train\_station\_count},
\texttt{poi\_healthcare}, \texttt{population\_total}\} to
\{\texttt{population\_total}\} alone, confirming that constraint-based methods
provide substantive independent votes when $n/p$ is large.
The $\tau = 5$ rule is therefore robust to reasonable perturbations.

\section{Citizen Preference Survey: Calibration Tables}
\label{app:survey}

\subsection*{Calibration Table 1: Mobility Usage Factors}

Age-stratified mobility mode usage probabilities derived from binary logistic regressions
fitted on the survey data (IBM SPSS), encoding the probability that a resident of a given
age group uses each mobility mode at a hub.
Values do not sum to 100\% as each mode is modelled independently.

{\setlength{\intextsep}{2pt}\setlength{\abovecaptionskip}{3pt}
\begin{table}[H]
\caption{Mobility usage probability by age group ($M$ matrix).}
\label{tab:mobility_factors}
\centering
\small
\begin{tabular}{lcccc}
\hline
\textbf{Age group} & \textbf{E-Scooter} & \textbf{Bicycle} & \textbf{Car} & \textbf{Other} \\
\hline
18--29 & 0.269 & 0.269 & 0.378 & 0.182 \\
30--49 & 0.182 & 0.289 & 0.401 & 0.231 \\
50--64 & 0.091 & 0.182 & 0.269 & 0.269 \\
$\geq$65 & 0.029 & 0.076 & 0.231 & 0.378 \\
\hline
\end{tabular}
\end{table}}

\subsection*{Calibration Table 2: Service Usage Ratings}

Mean survey ratings (1--5 scale) for five infrastructure service categories,
broken down by age group and reference area.
These values populate the preference matrix $W \in \mathbb{R}^{4 \times 5}$
used to compute the Companion Service Demand score for each candidate site.

{\setlength{\intextsep}{2pt}\setlength{\abovecaptionskip}{3pt}
\begin{table}[H]
\caption{Age-stratified infrastructure service ratings by reference area ($W$ matrix entries, normalised to 0--100\%).}
\label{tab:service_ratings}
\centering
\small
\begin{tabular}{llccc}
\hline
\textbf{Service Category} & \textbf{Age Group}
  & \textbf{Germany} & \textbf{Heilbronn City} & \textbf{Heilbronn Region} \\
\hline
\multirow{4}{*}{Cycling}
  & 18--29 & 64.5\% & 66.5\% & 62.5\% \\
  & 30--49 & 63.5\% & 64.0\% & 58.5\% \\
  & 50--64 & 61.0\% & 43.5\% & 50.3\% \\
  & $\geq$65 & 42.5\% & 47.5\% & 44.8\% \\
\hline
\multirow{4}{*}{Infrastructure Comfort \& Safety}
  & 18--29 & 73.3\% & 75.8\% & 72.3\% \\
  & 30--49 & 61.0\% & 69.0\% & 72.0\% \\
  & 50--64 & 60.3\% & 64.5\% & 66.5\% \\
  & $\geq$65 & 64.3\% & 69.8\% & 74.0\% \\
\hline
\multirow{4}{*}{Third-Party Convenience}
  & 18--29 & 37.3\% & 35.3\% & 37.5\% \\
  & 30--49 & 36.3\% & 38.0\% & 34.5\% \\
  & 50--64 & 29.0\% & 30.5\% & 28.5\% \\
  & $\geq$65 & 22.8\% & 27.3\% & 27.8\% \\
\hline
\multirow{4}{*}{E-Charging}
  & 18--29 & 51.0\% & 54.3\% & 58.3\% \\
  & 30--49 & 50.5\% & 57.3\% & 57.8\% \\
  & 50--64 & 51.5\% & 48.0\% & 35.8\% \\
  & $\geq$65 & 38.8\% & 51.8\% & 78.5\% \\
\hline
\multirow{4}{*}{Parcel Locker}
  & 18--29 & 65.0\% & 76.8\% & 65.0\% \\
  & 30--49 & 60.8\% & 57.8\% & 60.5\% \\
  & 50--64 & 52.5\% & 44.5\% & 40.0\% \\
  & $\geq$65 & 40.0\% & 41.8\% & 41.0\% \\
\hline
\end{tabular}
\end{table}}

\section{C0 vs C1 Feature Frequency Tables}
\label{app:cluster_freq}

Table~\ref{tab:cluster_freq} reports the occurrence ratios
of all 27 features as direct causal drivers, separately for C0 (core
high-activity, 29 units) and C1 (peripheral low-activity, 28 units).
Features are grouped by the five semantic blocks used in the pipeline.

\begin{table}[H]
\caption{Feature occurrence ratios for C0 (left, 29 cities) and C1 (right,
28 cities; Pforzheim excluded). Ratio = fraction of cities in which the
feature appears as a direct cause of \texttt{trip\_count}.}
\label{tab:cluster_freq}
\centering\small
\begin{minipage}[t]{0.48\linewidth}
\centering
\textbf{C0 -- core high-activity}\\[4pt]
\begin{tabular}{llcc}
\hline
\textbf{Block} & \textbf{Feature} & $n$ & \textbf{\%} \\
\hline
\multirow{7}{*}{POI}
  & Recreation POIs              & 13 & 45 \\
  & Education POIs               & 12 & 41 \\
  & Shopping POIs                & 10 & 34 \\
  & Food \& drink POIs           &  9 & 31 \\
  & Tourism \& culture POIs      &  9 & 31 \\
  & Healthcare POIs              &  7 & 24 \\
  & Services POIs                &  6 & 21 \\
\hline
\multirow{7}{*}{Urban form}
  & Street intersections         & 13 & 45 \\
  & Structured parking count     & 10 & 34 \\
  & Residential building count   & 10 & 34 \\
  & Building coverage (\%)       & 10 & 34 \\
  & High-rise count              & 10 & 34 \\
  & Flat terrain (\%)            &  8 & 28 \\
  & Commercial building count    &  6 & 21 \\
\hline
\multirow{8}{*}{Transportation}
  & Train station count          & 11 & 38 \\
  & Train station proximity (m)  & 10 & 34 \\
  & Bus stop count               &  9 & 31 \\
  & Cycling infrastructure (km)  &  8 & 28 \\
  & Slow road share (\%)         &  8 & 28 \\
  & Bus stop proximity (m)       &  5 & 17 \\
  & Bus route count              &  0 &  0 \\
  & Train route count            &  0 &  0 \\
\hline
\multirow{2}{*}{Population}
  & Population total             & 10 & 34 \\
  & Dormitory count              &  5 & 17 \\
\hline
\multirow{3}{*}{Mobility supply}
  & Scooter stations             &  5 & 17 \\
  & Parcel locker count          &  4 & 14 \\
  & Car-sharing stations         &  3 & 10 \\
\hline
\end{tabular}
\end{minipage}
\hfill
\begin{minipage}[t]{0.48\linewidth}
\centering
\textbf{C1 -- peripheral low-activity}\\[4pt]
\begin{tabular}{llcc}
\hline
\textbf{Block} & \textbf{Feature} & $n$ & \textbf{\%} \\
\hline
\multirow{7}{*}{POI}
  & Recreation POIs              & 10 & 36 \\
  & Education POIs               &  9 & 32 \\
  & Tourism \& culture POIs      &  7 & 25 \\
  & Shopping POIs                &  7 & 25 \\
  & Healthcare POIs              &  6 & 21 \\
  & Services POIs                &  5 & 18 \\
  & Food \& drink POIs           &  2 &  7 \\
\hline
\multirow{7}{*}{Urban form}
  & Building coverage (\%)       &  9 & 32 \\
  & High-rise count              &  9 & 32 \\
  & Structured parking count     &  9 & 32 \\
  & Commercial building count    &  8 & 29 \\
  & Flat terrain (\%)            &  7 & 25 \\
  & Street intersections         &  5 & 18 \\
  & Residential building count   &  5 & 18 \\
\hline
\multirow{8}{*}{Transportation}
  & Slow road share (\%)         &  8 & 29 \\
  & Train station proximity (m)  &  7 & 25 \\
  & Bus stop count               &  7 & 25 \\
  & Train station count          &  6 & 21 \\
  & Train route count            &  4 & 14 \\
  & Cycling infrastructure (km)  &  3 & 11 \\
  & Bus stop proximity (m)       &  2 &  7 \\
  & Bus route count              &  2 &  7 \\
\hline
\multirow{2}{*}{Population}
  & Population total             &  2 &  7 \\
  & Dormitory count              &  1 &  4 \\
\hline
\multirow{3}{*}{Mobility supply}
  & Car-sharing stations         &  6 & 21 \\
  & Scooter stations             &  5 & 18 \\
  & Parcel locker count          &  3 & 11 \\
\hline
\end{tabular}
\end{minipage}
\end{table}

\section{Overview: All City-Cluster Analysis Units}
\label{app:all_units}

Table~\ref{tab:all_units} summarises all 57 city-cluster analysis units across 29 cities.
$N_h$ = number of hotspots in the cluster.
High-conf = consensus edges with high bootstrap support; Med-conf = moderate support;
`---' = no causes identified at that confidence level.

{\scriptsize
\renewcommand{\_}{\nobreak\textunderscore\nobreak}%
\newcommand{\rowsep}{\arrayrulecolor{gray!50}\hline\arrayrulecolor{black}}%
\setlength{\tabcolsep}{3pt}
\begin{longtable}{llr >{\raggedright\arraybackslash}p{4.8cm} >{\raggedright\arraybackslash}p{6.8cm}}
\caption{All city-cluster analysis units (29 cities, 57 clusters).
High-conf / Med-conf: consensus edges in $\geq$2 methods with high / moderate bootstrap support;
`---' = none identified.}
\label{tab:all_units}\\
\toprule
City & C & $N_h$ & High-conf causes & Med-conf causes \\
\midrule
\endfirsthead
\multicolumn{5}{l}{\textit{(continued from previous page)}}\\
\toprule
City & C & $N_h$ & High-conf causes & Med-conf causes \\
\midrule
\endhead
\midrule
\multicolumn{5}{r}{\textit{continued on next page}}\\
\endfoot
\bottomrule
\endlastfoot
Berlin & C0 & 919 & dormitory\_count, poi\_food\_drink & scooter\_stations, train\_station\_count, bus\_stop\_count, nearest\_train\_station\_meter, poi\_education \\
\rowsep
Berlin & C1 & 101 & --- & nearest\_train\_station\_meter, building\_coverage\_pct, highrise\_count \\
\rowsep
B\"oblingen & C0 &  64 & structured\_parking\_count, poi\_services & cycling\_infra\_km, train\_station\_count, pct\_slow\_roads, poi\_shopping, poi\_education\ldots \\
\rowsep
B\"oblingen & C1 &  56 & --- & pct\_flat\_terrain, bld\_commercial\_count \\
\rowsep
Bonn & C0 & 282 & --- & poi\_shopping, poi\_recreation, poi\_tourism\_culture, building\_coverage\_pct, street\_intersections\ldots \\
\rowsep
Bonn & C1 & 189 & --- & poi\_recreation, poi\_healthcare, poi\_services, nearest\_train\_station\_meter, poi\_shopping \\
\rowsep
Braunschweig & C0 & 279 & street\_intersections & bld\_residential\_count, poi\_food\_drink \\
\rowsep
Braunschweig & C1 &  93 & --- & poi\_recreation, poi\_services, building\_coverage\_pct, highrise\_count, bus\_stop\_count\ldots \\
\rowsep
Chemnitz & C0 & 279 & poi\_recreation, population\_total & nearest\_bus\_stop\_meter, bus\_stop\_count \\
\rowsep
Chemnitz & C1 & 129 & --- & poi\_recreation, poi\_education, poi\_healthcare, bus\_stop\_count, street\_intersections\ldots \\
\rowsep
Darmstadt & C0 &  54 & poi\_tourism\_culture, structured\_parking\_count, dormitory\_count, poi\_education, carsharing\_stations & nearest\_bus\_stop\_meter, pct\_slow\_roads, poi\_recreation, bld\_commercial\_count, highrise\_count\ldots \\
\rowsep
Darmstadt & C1 & 134 & bld\_residential\_count, poi\_tourism\_culture & poi\_education, poi\_healthcare, pct\_flat\_terrain, pct\_slow\_roads, poi\_shopping \\
\rowsep
Dortmund & C0 & 118 & poi\_healthcare & poi\_tourism\_culture, poi\_recreation, bld\_residential\_count, street\_intersections, pct\_slow\_roads\ldots \\
\rowsep
Dortmund & C1 &  23 & --- & cycling\_infra\_km, train\_station\_count, pct\_slow\_roads, structured\_parking\_count, highrise\_count\ldots \\
\rowsep
D\"usseldorf & C0 & 259 & highrise\_count, train\_station\_count, building\_coverage\_pct, poi\_recreation & poi\_services, bld\_commercial\_count, bld\_residential\_count, street\_intersections, cycling\_infra\_km\ldots \\
\rowsep
D\"usseldorf & C1 & 123 & poi\_education & train\_station\_count, street\_intersections, poi\_recreation, highrise\_count, carsharing\_stations \\
\rowsep
Erfurt & C0 & 114 & highrise\_count & cycling\_infra\_km, poi\_shopping, poi\_recreation, street\_intersections \\
\rowsep
Erfurt & C1 &  11 & poi\_education & --- \\
\rowsep
Erlangen & C0 &  31 & poi\_services & poi\_tourism\_culture, poi\_food\_drink, highrise\_count, bld\_residential\_count, poi\_education\ldots \\
\rowsep
Erlangen & C1 & 117 & --- & --- \\
\rowsep
Friedrichshafen & C0 & 107& --- & nearest\_train\_station\_meter, poi\_recreation, poi\_food\_drink, poi\_education, bld\_commercial\_count\ldots \\
\rowsep
Friedrichshafen & C1 &  33& poi\_recreation, street\_intersections & poi\_tourism\_culture, train\_route\_count, pct\_slow\_roads, nearest\_train\_station\_meter, poi\_food\_drink\ldots \\
\rowsep
Hamburg & C0 & 1169& train\_station\_count, population\_total & cycling\_infra\_km, poi\_shopping, poi\_food\_drink, poi\_healthcare, poi\_tourism\_culture \\
\rowsep
Hamburg & C1 & 1470& --- & --- \\
\rowsep
Heilbronn & C0 & 164 & street\_intersections & population\_total, parcel\_locker\_count, bus\_stop\_count, poi\_recreation, poi\_food\_drink\ldots \\
\rowsep
Heilbronn & C1 &  42 & nearest\_train\_station\_meter & train\_route\_count, train\_station\_count, bus\_stop\_count, poi\_tourism\_culture, poi\_shopping\ldots \\
\rowsep
Ingolstadt & C0 & 192 & poi\_healthcare & poi\_education, building\_coverage\_pct, highrise\_count, pct\_flat\_terrain, parcel\_locker\_count\ldots \\
\rowsep
Ingolstadt & C1 & 115 & bld\_commercial\_count & poi\_healthcare, poi\_recreation, structured\_parking\_count \\
\rowsep
Jena & C0 &  80 & --- & train\_station\_count, poi\_healthcare, cycling\_infra\_km, bus\_stop\_count, poi\_education\ldots \\
\rowsep
Jena & C1 &  16 & building\_coverage\_pct, highrise\_count & --- \\
\rowsep
Kaiserslautern & C0 & 155& --- & nearest\_train\_station\_meter, poi\_recreation \\
\rowsep
Kaiserslautern & C1 &  38& pct\_flat\_terrain & pct\_slow\_roads, poi\_food\_drink, poi\_education, structured\_parking\_count, train\_route\_count\ldots \\
\rowsep
Karlsruhe & C0 & 315 & carsharing\_stations, bus\_stop\_count & structured\_parking\_count, train\_station\_count, poi\_tourism\_culture, poi\_shopping, poi\_healthcare\ldots \\
\rowsep
Karlsruhe & C1 & 135 & --- & poi\_tourism\_culture, poi\_recreation, poi\_education, structured\_parking\_count, parcel\_locker\_count\ldots \\
\rowsep
Kiel & C0 & 168 & structured\_parking\_count, scooter\_stations & nearest\_train\_station\_meter, pct\_slow\_roads, poi\_healthcare, poi\_education, street\_intersections\ldots \\
\rowsep
Kiel & C1 &  83 & --- & poi\_shopping, poi\_education, structured\_parking\_count, bld\_residential\_count, carsharing\_stations\ldots \\
\rowsep
K\"oln & C0 & 413 & structured\_parking\_count & poi\_services, pct\_slow\_roads, train\_station\_count, highrise\_count \\
\rowsep
K\"oln & C1 & 523 & --- & --- \\
\rowsep
Ludwigsburg & C0 & 135 & --- & nearest\_train\_station\_meter, poi\_shopping \\
\rowsep
Ludwigsburg & C1 &  19 & bld\_commercial\_count & --- \\
\rowsep
Mannheim & C0 & 280 & building\_coverage\_pct, highrise\_count, carsharing\_stations, poi\_education, dormitory\_count & poi\_recreation, pct\_flat\_terrain, street\_intersections, cycling\_infra\_km, poi\_shopping \\
\rowsep
Mannheim & C1 & 177 & --- & highrise\_count \\
\rowsep
Pforzheim & C0 & 141 & street\_intersections, poi\_shopping & cycling\_infra\_km, nearest\_train\_station\_meter, poi\_tourism\_culture, poi\_education, bld\_residential\_count\ldots \\
\rowsep
Reutlingen & C0 &  92 & --- & nearest\_train\_station\_meter, poi\_education, poi\_services, pct\_flat\_terrain, building\_coverage\_pct\ldots \\
\rowsep
Reutlingen & C1 & 171 & poi\_tourism\_culture & bus\_stop\_count, pct\_slow\_roads, nearest\_train\_station\_meter, poi\_recreation, poi\_education\ldots \\
\rowsep
Rostock & C0 &  38 & --- & nearest\_bus\_stop\_meter, train\_station\_count, poi\_tourism\_culture, pct\_flat\_terrain, poi\_shopping\ldots \\
\rowsep
Rostock & C1 & 158 & --- & train\_station\_count, bld\_commercial\_count, building\_coverage\_pct, nearest\_train\_station\_meter \\
\rowsep
Saarbr\"ucken & C0 & 216 & --- & bld\_residential\_count \\
\rowsep
Saarbr\"ucken & C1 & 147 & poi\_recreation & train\_station\_count, structured\_parking\_count, highrise\_count, cycling\_infra\_km, poi\_services\ldots \\
\rowsep
Stuttgart & C0 & 411 & poi\_education, building\_coverage\_pct, pct\_flat\_terrain & bld\_residential\_count, train\_station\_count, bus\_stop\_count, poi\_healthcare, structured\_parking\_count\ldots \\
\rowsep
Stuttgart & C1 & 492 & building\_coverage\_pct, pct\_flat\_terrain, population\_total & poi\_education, bld\_residential\_count, train\_station\_count \\
\rowsep
T\"ubingen & C0 &  38 & bld\_commercial\_count & cycling\_infra\_km, pct\_slow\_roads, poi\_shopping, building\_coverage\_pct, structured\_parking\_count\ldots \\
\rowsep
T\"ubingen & C1 & 141 & --- & bus\_stop\_count, building\_coverage\_pct, nearest\_train\_station\_meter \\
\rowsep
Wolfsburg & C0 & 196 & building\_coverage\_pct & nearest\_train\_station\_meter, highrise\_count \\
\rowsep
Wolfsburg & C1 & 116 & pct\_slow\_roads, bus\_stop\_count, poi\_recreation & cycling\_infra\_km, poi\_tourism\_culture, poi\_shopping, highrise\_count\ldots \\
\rowsep
Wuppertal & C0 & 304 & bld\_residential\_count, population\_total & bus\_stop\_count, nearest\_train\_station\_meter, poi\_recreation, poi\_food\_drink, scooter\_stations \\
\rowsep
Wuppertal & C1 &  51 & --- & poi\_healthcare, bld\_residential\_count, bld\_commercial\_count, scooter\_stations \\
\rowsep
\end{longtable}
}

\bibliographystyle{elsarticle-harv}
\bibliography{references}
\end{document}